\documentclass[12pt]{article}
\usepackage{chetnew}
\usepackage{amssymb,amsmath,amsfonts,manfnt,graphicx}
\usepackage{textcomp,braket,slashed,tikz,tikz-cd,pst-plot}
\usepackage[utf8]{inputenc}
\usepackage{mathtools}

\usetikzlibrary{arrows,decorations.markings,shapes.geometric,decorations.pathmorphing}
\tikzset{snake it/.style={decorate, decoration=snake}}
\usetikzlibrary{arrows}

\newcommand{\overbar}[1]{\mkern 1.5mu\overline{\mkern-1.5mu#1\mkern-1.5mu}\mkern 1.5mu}

\def\one{{\,\hbox{1\kern-.8mm l}}}

\def\d{{\partial}}

\def\IT{{\mathbb T}}

\def\IH{{\mathbb H}}
\def\half{{\frac{1}{2}}}
\def\q{{\mathfrak q}}

\newcommand{\midarrow}{\tikz \draw[-triangle 90] (0,0) -- +(.1,0);}
\newcommand{\miduno}{\tikz \draw[-triangle 90, rotate=45](0,0) -- +(.1,0);}
\newcommand{\middue}{\tikz \draw[-triangle 90, rotate=-45](0,0) -- +(.1,0);}
\newcommand{\midup}{\tikz \draw[-triangle 90, rotate=90](0,0) -- +(.1,0);}
\newcommand{\middiag}{\tikz \draw[-triangle 90, rotate=30](0,0) -- +(0.01,0.01);}

\newcommand{\Tr}{\mathrm{Tr}}

\newcommand{\CC}{\mathcal{C}}

\newcommand{\QQ}{\mathcal{Q}}

\newcommand{\NN}{\mathcal N}

\newcommand{\II}{\mathcal I}
\newcommand{\OO}{\mathcal O}

\def\beq{\begin{equation}}
\def\eeq{\end{equation}}
\newcommand{\bea}{\begin{eqnarray}}
\newcommand{\eea}{\end{eqnarray}}
\def\bal{\begin{align}}
\def\eal{\end{align}}
\newcommand\Qeq{\stackrel{\mathclap{\normalfont\mbox{?}}}{=}}

\preprint{CCTP-2020-10\hfill QMUL-PH-20-26\\DCPT-20/13 \hfill
  DESY-20-145\\ ITCP-IPP-2020/10}

\title{\vspace{-1.5cm}(Mis-)Matching Type-B Anomalies  on the  Higgs Branch
}

\author{V.~Niarchos\;$^{a,b,\clubsuit}$, C.~Papageorgakis\;$^{c,\diamondsuit}$, A.~Pini\;$^{d,e,\spadesuit}$ and E.~Pomoni\;$^{e,\heartsuit}$}

\affiliation{$^a$ ITCP \& CCTP, Department of Physics,\\
University of Crete, 71003 Heraklion, Greece
\vspace{0.3cm} $ $\\

$^b$Department of Mathematical Sciences and Centre for Particle Theory,\\ 
Durham University, Durham DH1 3LE, UK \vspace{0.3cm} $ $ \\

$^c$CRST and School of Physics and Astronomy\\ Queen Mary University of London, London E1 4NS, UK \vspace{0.3cm} $ $ \\

$^d$I.N.F.N.- Sezione di Torino, Via Pietro Giuria 1, 10125 Torino, Italy \vspace{0.3cm} $ $\\
$^e$DESY, Theory Group, Notkestrasse 85, Building 2a, 22607 Hamburg, Germany \\

\emails{$^\clubsuit$niarchos@physics.uoc.gr, $^\diamondsuit$c.papageorgakis@qmul.ac.uk, $^\spadesuit$alessandro.pini@desy.de, $^\heartsuit$elli.pomoni@desy.de}}

\abstract{Building on \cite{Niarchos:2019onf}, we uncover new properties of type-B conformal anomalies for Coulomb-branch operators in continuous families of 4D $\mathcal{N}=2$ SCFTs. We study a large class of such anomalies on the Higgs branch, where conformal symmetry is spontaneously broken, and compare them with their counterpart in the CFT phase. In Lagrangian theories, the non-perturbative matching of the anomalies can be determined with a weak coupling Feynman diagram computation involving massive multi-loop banana integrals. We extract the part corresponding to the anomalies of interest. Our calculations support the general conjecture that the Coulomb-branch type-B conformal anomalies always match on the Higgs branch. On the other hand, we argue that the potential mismatch of anomalies implies the existence of a second covariantly constant metric on the conformal manifold (other than the Zamolodchikov metric), which would impose restrictions on its holonomy group.  \bigskip
}

\date{
}

\begin{document}
\maketitle

\hypersetup{pageanchor=true}

\setcounter{tocdepth}{2}

\toc

\section{Introduction}
\label{intro}

Anomaly matching is a powerful tool in Quantum Field Theory. It provides a useful window into non-perturbative physics and finds applications in many contexts. One of the most popular examples involves chiral anomalies, which match across different phases due to 't Hooft's argument \cite{tHooft:1979rat}. In this work we will be interested in a different type of anomaly, one associated with the breaking of conformal symmetry.

Conformal field theories (CFTs) in even spacetime dimensions  exhibit two classes of conformal anomalies \cite{Deser:1993yx}. The first is the so-called type-A class and is well-studied. These Weyl anomalies do not introduce a scale. At the level of the generating functional of correlation functions, they are expressed  in terms of topological invariants and must match across different phases of the theory \cite{Schwimmer:2010za}. In this sense, type-A conformal anomalies are akin to chiral anomalies. The coefficient $a$, multiplying the Euler-density term in the generating functional of 4D CFTs, is a well-known example of a type-A anomaly. 

There is also a type-B class of conformal anomalies. These are associated with the presence of logarithmic divergences in specific correlation functions and the introduction of a corresponding scale. In contrast to type-A, these anomalies are generically not expected to match across different phases, but there are exceptions. The coefficient $c$ of the energy-momentum tensor two-point function in 4D CFTs, is a well-known example of a type-B conformal anomaly. The study of type-B anomalies across different phases of a CFT will be the main focus of this work.

More specifically, whenever an (even-) D-dimensional theory possesses operators with integer scaling dimension $\Delta=\frac{\mathrm D}{2}+n$, $n\in {\mathbb Z}_{\geq 0}$, there is a corresponding type-B Weyl anomaly. For two operators $\OO_I$, $\OO_J$ with common scaling dimension $\Delta=\frac{\mathrm D}{2}+n$ the logarithmically divergent part of the two-point function
\beq
\label{introaa}
\langle \OO_I(p) \OO_J(-p)\rangle \simeq (-1)^{n+1} \frac{\pi^2 G_{IJ}}{2^{2n} \Gamma(n+1) \Gamma(\frac{D}{2}+n)} p^{2n} \log\left( \frac{p^2}{\mu^2} \right)
\eeq
introduces the momentum scale $\mu$. The type-B anomaly is then expressed in terms of the two-point function coefficient $G_{IJ}$ \cite{Osborn:1991gm,Osborn:1993cr,Petkou:1999fv}. In addition to the energy-momentum tensor, CFTs (especially supersymmetric ones) frequently possess several operators that have integer scaling dimension, and all of these operators lead to corresponding type-B Weyl anomalies.

In particular, superconformal field theories (SCFTs) with $\NN=2$ supersymmetry in four dimensions typically contain scalar superconformal primary operators in protected $\half$-BPS superconformal multiplets, whose scaling dimension $\Delta= |r|$ is an integer, given in terms of the operator $U(1)_r$ R-charge, $r$.
They are usually referred to as Coulomb-branch operators (CBOs) and form a ring under the operator product expansion (OPE)---the Coulomb-branch chiral ring. The two-point function coefficient of CBOs, $G_{I\bar J}$, defines a natural Hermitian metric on the corresponding vector space of operators. This metric is  intimately related to the $S^4$ partition function of the theory \cite{Gerchkovitz:2014gta,Gerchkovitz:2016gxx}, and in many cases can be extracted using methods of supersymmetric localisation \cite{Pestun:2007rz} or the AGT relation \cite{Alday:2009aq}.

In this paper, we explore whether type-B Weyl anomalies for CBOs match in different phases of
4D $\NN=2$ SCFTs with a non-trivial $\NN=2$ conformal manifold. Continuing work initiated in \cite{Niarchos:2019onf}, we analyse the properties of such type-B anomalies on the Higgs branch where the conformal symmetry is spontaneously broken.  In \cite{Niarchos:2019onf,Andriolo:2022lcb} we argued that CBO type-B anomalies continue to be covariantly constant (with respect to the appropriate holomorphic connection) along the Higgs branch, just as in the conformal phase. That is
\beq
\label{eq:CovariantlyConstant}
\nabla G_{I\bar J}^{\rm CFT}(g) =0
   \quad \mbox{and} \quad  
\nabla  G_{I\bar J}^\IH(g) =0     ~, 
\eeq
 as will be reviewed in  Sec.~\ref{Sec:Review}. $g$ denotes collectively the exactly marginal couplings that parametrise the conformal manifold. Although this property is constraining, it does not provide a verdict on whether the Higgs-branch value of the anomaly equals the value in the conformally symmetric phase. To settle this question one needs to perform an independent computation at any one point of the conformal manifold  $g^*$:
 \beq
 \label{eq:matching}
 G_{I\bar J}^\IH(g^*) \Qeq G_{I\bar J}^{\rm CFT}(g^*)  \,  .
 \eeq
Therefore, for theories with a Lagrangian description the tree-level, $g^*=0$, computation is sufficient to establish the (mis-)matching non-perturbatively.

In \cite{Niarchos:2019onf} we worked out two Lagrangian examples of type-B CBO anomalies. In the first example of the 4D $\NN=2$ superconformal QCD (SCQCD), the type-B anomalies for scaling dimension $\Delta=2$ CBOs were found to match in the two phases (CFT and Higgs branch).
In the second example of a 4D $\NN=2$ circular quiver, \cite{Niarchos:2019onf} initially reported that the corresponding type-B anomalies generically do not match.
This statement was based on an erroneous factor in the relevant perturbative computation. Accounting for this factor yields matching anomalies also for this case.

These observations favour a general picture of matching for type-B CBO anomalies on the Higgs branch of 4D $\mathcal N=2$ SCFTs, 
which is formulated in the form of two conjectures in Sec.~\ref{conjectures}. More specifically, we propose: I) that type-B CBO anomalies match when a potential non-trivial Coulomb-branch chiral ring in the extreme low-energy theory on the Higgs branch is properly taken into account, and II) that the chiral ring OPE preserves the matching of type-B CBO anomalies on the Higgs branch. Favourable evidence for both conjectures is presented in Secs \ref{sec:N2SCQCD} and \ref{sec:quiver}.

In Sec.~\ref{sec:N2SCQCD} we analyse the anomalies of arbitrary CBOs on the Higgs branch of the 4D $\NN=2$ SCQCD theory at leading order in the Yang--Mills coupling. The computation involves Feynman diagrams  with massive propagators leading to multi-loop `banana integrals' (similar to the `banana integrals' that appear in QCD). Although, these integrals are notoriously hard to evaluate exactly, we show that the anomaly of interest in the Higgs branch is a very special, simple part of the full integral that matches the anomaly in the conformal phase. 

In the example of the $\NN=2$ circular quiver of Sec.\ \ref{sec:quiver}, the generic type-B CBO anomalies are also found to match along the Higgs branch. In particular, one linear combination of the  $\NN=2$ vector multiplets is special as it remains massless all the way to the extreme infrared (IR). In accordance with the conjectures of Sec.~\ref{conjectures}, an  extra contribution to the anomalies from this sector is crucial for the corresponding type-B CBO anomaly matching in this context. A highly non-trivial, Feynman diagram computation in Sec.~\ref{sec:quiver} confirms that this is indeed the case at leading order. Although not needed for the matching, computations that go beyond leading order in the CFT phase (with the help of supersymmetric localisation) are presented in Sec.~\ref{localisation}.

In Sec.~\ref{holonomy} we ask what would happen if the CBO anomalies of 4D $\mathcal N=2$ SCFTs do not match on the Higgs branch and investigate potential implications of a type-B anomaly mismatch for the conformal manifold. The conformal manifold is the set of all possible values of exactly marginal coupling constants of a SCFT. As such it is endowed with the structure of Riemannian geometry \cite{Zamolodchikov:1986gt,Kutasov:1988xb,Ranganathan:1993vj}, a complex structure and a K\"ahler metric \cite{Seiberg:1988pf,Asnin:2009xx}.  For 4D $\NN=2$ SCFTs the $\Delta=2$ CBOs are in the same superconformal multiplet as the exactly marginal operators, thus the Zamolodchikov metric \cite{Zamolodchikov:1986gt} is identical to the Hermitian metric on the holomorphic vector bundle of $\Delta=2$ CBOs $\OO_I$, $\overbar \OO_J$, with components $G_{I\bar J}$. It follows from \eqref{eq:CovariantlyConstant} that if the anomalies of a set of CBOs do not match on the Higgs branch, then there are two sets of metrics on the corresponding holomorphic vector bundle that are compatible with the same connection. If the two metrics are genuinely different---namely if they are not equal up to a constant factor---then their existence constrains the holonomy of the connection on the vector bundle, and in the case of $\Delta=2$ CBOs, leads to a novel constraint on the holonomy of $\NN=2$ superconformal manifolds. Finally, in Sec.~\ref{sec:mismatch} we point out that aspects of a potential mismatch could be quantified in terms of a scalar quantity. This quantity is constant on the superconformal manifold, constant along the RG flow on the Higgs branch, and independent of the normalisation of the CBOs.

\vskip .3cm
\noindent {\bf Note added (December 2024):} Previous versions of this work relied on an observed mismatch between the CFT- and Higgs-phase type-B anomalies of the $N$-noded $\mathcal N=2$ circular-quiver theory at finite $N$ and at tree level. We thank A.~Schwimmer and S.~Theisen for pointing out a computational error in our previous calculations. The correct perturbative calculation implies that the anomalies match for all $N$. As a result, there currently is no known example where the type-B anomalies do not match in different phases of a given theory. In light of more recent results that have clarified the matching of type-B anomalies \cite{Schwimmer:2023nzk, Schwimmer:2024vxw}, some of the conjectures proposed 
in Sec.~\ref{conjectures} of this paper have been updated accordingly. The computations of Secs~\ref{sec:N2SCQCD}, \ref{sec:quiver}, \ref{localisation} remain relevant. The 
discussion of Secs~\ref{holonomy}, \ref{sec:mismatch} explore the implications of a potential anomaly mismatch, and as such, we believe that it is still useful.

\section{General Structure of Type-B Conformal Anomalies}
\label{Sec:Review}

\subsection{Summary of Key Properties and Open Questions}

A very useful aspect of chiral anomalies is 't Hooft anomaly matching. This is the statement that chiral anomalies match across different scales along an RG flow. It is natural to ask if conformal anomalies share the same property. More specifically, one can ask if conformal anomalies match across different phases of a theory, namely across different scales in an RG flow generated by a vacuum expectation value (VEV).\footnote{The very interesting question of matching across RG flows generated by relevant deformations of the theory will not be considered in this paper. Arguments in favour of type-A anomaly matching for relevant deformations have been given in the literature using conformal compensator fields, see e.g.\ \cite{Komargodski:2011vj}. For type-B anomalies this question is much harder and essentially unexplored. We will make brief comments on this case in Sec.~\ref{sec:mismatch}.} Type-A Weyl anomalies---like the coefficient $a$ in 4D SCFTs---match across different phases of a theory \cite{Schwimmer:2010za}. The general arguments used to reach this conclusion do not rely on supersymmetry.

The case of type-B anomalies is more involved. The following features suggest that type-B anomalies are generically {\it not} expected to match across different phases:
\begin{itemize}

\item[$(a)$] Type-B anomalies can depend non-trivially on continuous exactly marginal couplings on a conformal manifold. In contrast, the Wess--Zumino consistency conditions can be used to show that type-A anomalies do not have such a dependence and are usually expressed in terms of a few discrete data of the theory \cite{Nakayama:2017oye}  (e.g.\ the rank of the gauge group in a gauge theory).

\item[$(b)$] As we have already noted, in a phase that has the full conformal symmetry, type-B anomalies are directly related to two-point function coefficients. The relation with the two-point functions is absent in phases where the conformal symmetry is spontaneously broken. In such cases, the analytic structure of correlation functions changes and the dilaton---the Goldstone boson for the spontaneous breaking of conformal symmetry---plays a crucial role in the way the anomaly manifests itself. 

\end{itemize}

It will be useful to recall some of the details entering in item $(b)$ and how they affect the definition of type-B conformal anomalies in phases with spontaneously broken conformal symmetry. For a detailed discussion of the material presented here we refer the reader to \cite{Niarchos:2019onf} and references therein. In what follows, we focus for concreteness on the case of four-dimensional CFTs. Let us denote the three-point function of the energy-momentum tensor $T_{\mu\nu}$ with two operators $\OO$ and $\bar \OO$, that have the same integer scaling dimension $\Delta=2+n$ $(n=0,1,2,\ldots)$,  as 
\beq
\label{wardaa}
\Gamma_{\mu\nu}^{(3)}(q,k_1,k_2) = \langle T_{\mu\nu}(q) \OO(k_1) \bar \OO(k_2) \rangle
~.
\eeq
Let us also denote the two-point function of the operators $\OO,\bar \OO$ as
\beq
\label{wardab}
\Gamma^{(2)}(k^2) = \langle \OO(k) \bar \OO(-k) \rangle
~.
\eeq
The tensor structure of $\Gamma^{(3)}_{\mu\nu}$ can be expressed in the form \cite{Schwimmer:2010za}
\beq
\label{wardac}
\Gamma_{\mu\nu}^{(3)} = \bar A \eta_{\mu\nu} + B q_{\mu} q_{\nu} +C(q_\mu r_\nu + q_\nu r_\mu) +D r_\mu r_\nu
~,
\eeq
where $\bar A, B, C, D$ depend on the Lorentz invariants $q^2, k_1^2, k_2^2$ and $r\equiv k_1-k_2$. It is also convenient to define the combination
\beq
\label{wardad}
A \equiv \bar A -\frac{1}{4} \Big( \Gamma^{(2)}(k_1^2) + \Gamma^{(2)}(k_2^2) \Big)
~.
\eeq

In a conformally symmetric phase one can show (see \cite{Niarchos:2019onf} for a review of the relevant details) that the diffeomorphism and Weyl Ward identities lead, respectively, to the mutually contradicting relations
\beq
\label{wardae}
A=0~, ~~ 4A = - G^{\rm CFT} k^{2n}
\eeq
in the kinematic regime $q^2=0$, $k_1^2=k_2^2 = k^2$. $G^{\rm CFT}$ is the two-point function coefficient in $\Gamma^{(2)}$. As advertised, the type-B anomaly is connected directly to the two-point function coefficient.

In phases with spontaneously broken conformal symmetry there is a different source for the type-B anomaly. In the kinematical regime $q^2 \to 0$, $k_1^2=k_2^2=k^2 \to 0$, one can argue that the dilaton can contribute a pole to the $B$ coefficient in $\Gamma^{(3)}_{\mu\nu}$, which leads to a non-vanishing term of the form
\beq
\label{wardaf}
\lim_{q^2 \to 0} q^2 B \supset G^{(\rm dil)} k^{2n}
\eeq
at order $k^{2n}$ in the low $k$-momentum  expansion of $B$. The non-vanishing coefficient $G^{(\rm dil)}$ contributes to the type-B anomaly because the diffeomorphism and Weyl Ward identities lead, respectively, to the following mutually contradicting relations
\beq
\label{wardag}
A+G^{(\rm dil)} k^{2n}=0~, ~~
4A+ G^{(\rm dil)} k^{2n}=-\left[ k^2 \frac{\d \Gamma^{(2)}}{\d k^2} - n \Gamma^{(2)}(k^2) \right]_{k^{2n}}
~.
\eeq
The notation $[\cdots]_{k^{2n}}$ denotes the $k^{2n}$ term in the low-momentum expansion of the quantity inside the parenthesis. When this quantity is analytic around $k^2=0$ the RHS on the second equation in \eqref{wardag} does not contribute and the clash between the two Ward identities is accounted for completely by the coefficient $G^{(\rm dil)}$ in the $B$ term of $\Gamma^{(3)}$. This is the case when the operators $\OO, \bar \OO$ only carry contributions from massive degrees of freedom. These degrees of freedom are lifted completely in the extreme IR of the theory in the broken phase. 

In contrast, when the operators $\OO, \bar \OO$ survive in the IR their two-point function $\Gamma^{(2)}$ can exhibit a logarithmic dependence on $k^2$ (similar to the logarithmic dependence that the two-point function exhibits in the ultraviolet (UV), which is dominated by the physics of the unbroken phase). Hence, if there is a piece in the IR two-point function that behaves as
\beq
\label{wardai}
\Gamma^{(2)}(k^2) \sim G^{(\rm IR)}k^{2n} \log \left( \frac{k^2}{\mu^2} \right)
\eeq
the RHS of the second equation in \eqref{wardag} does not vanish. Instead, from \eqref{wardag} one obtains
\beq
\label{wardaj}
A+G^{(\rm dil)} k^{2n}=0~, ~~
4A+\left(G^{(\rm dil)} +  G^{(\rm IR)}\right) k^{2n}= 0
~.
\eeq
There is still a type-B Weyl anomaly, but now it receives contributions {\it both} from {\it massive} degrees of freedom (the $G^{(\rm dil)}$ part) and {\it massless} degrees of freedom (the $G^{(\rm IR)}$ part). This anomaly manifests itself as a contact term in the appropriate low-momentum limit of the three-point function $\langle T^\mu_\mu(q) \OO(k_1) \bar \OO(k_2)\rangle$, which scales as $k^{2n}$ in momentum space with an overall coefficient 
\beq
G^{\rm SSB} = G^{(\rm dil)} + G^{(\rm IR)} \,.
\eeq
 $G^{\rm SSB}$ defines the type-B anomaly in a phase of spontaneously broken conformal symmetry. The potential contribution of massless IR degrees of freedom in $G^{\rm SSB}$ was not appreciated in \cite{Niarchos:2019onf}. We will see that it plays a crucial role in the computations of Sec.~\ref{sec:quiver} and in the context of Conjecture Ib in the upcoming Sec.~\ref{conjectures}.

Let us summarise a proper definition of the type-B anomaly coefficients, which recovers the above results in all phases of a CFT (with or without spontaneously broken conformal symmetry). We present the definition in arbitrary even spacetime dimension D. Consider two operators $\OO_I$, $\OO_J$ with common scaling dimension $\Delta = \frac{\rm D}{2}+n$. The type-B anomaly of interest is expressed in momentum space as the coefficient
\beq
\label{introba}
 G^{\rm SSB}_{IJ} = (-1)^n \frac{2^{2n}\Gamma(n+1)\Gamma(\frac{\rm D}{2}+n)}{\pi^2 (n!)^2} \lim_{p_1\to 0} \lim_{p_2,p_3\to 0} \left[ \frac{d^n}{d p_2^n} \frac{d^n}{dp_3^n} \left( \langle T(p_1) \OO_I(p_2) \OO_J(p_3) \rangle \right) \right]
~,
\eeq
where $T=T^\mu_\mu$ is the trace of the energy-momentum tensor. In the symmetric phase, the definition \eqref{introba} recovers the two-point function coefficient $G^{\rm CFT}_{IJ}$.\footnote{For the explicit evaluation of the three-point function $\Gamma^{(3)}$ in the unbroken phase see e.g. \cite{Bzowski:2018fql,Schwimmer:2019efk}. In the broken phase this definition appeared in \cite{Niarchos:2019onf}, and is already implicit in Eq.~\eqref{wardaf} in the context of the Ward identities with the $k$ dependence dictated by dimensional analysis.}

It is interesting to ask:
\begin{itemize}
\item[$(i)$] Are there special mechanisms that force type-B anomalies to match across different phases contrary to the above expectations? This is particularly interesting when the anomalies depend non-trivially on continuous couplings. 
\item[$(ii)$] If the anomalies do not match, can one identify the origin of the mismatch and the relevant physics of the broken phase? In particular, one would like to compute the anomalies non-perturbatively. When the anomalies depend non-trivially on continuous couplings they can interpolate between different behaviours at weak and strong coupling and one would like to know what effects determine this interpolation. 
\end{itemize}

An example of type-B anomaly matching envisaged in item $(i)$ occurs in the case of the $c$-anomaly in 4D SCFTs. Since supersymmetry relates the $c$-anomaly to chiral anomalies, it is expected that $c$ matches across different phases by standard 't Hooft anomaly matching. A less trivial mechanism of type-B anomaly matching occurs in the Higgs branch of 4D $\NN=2$ SCFTs \cite{Niarchos:2019onf}. This mechanism is the main focus in the remainder of this paper.

\subsection{Type-B Anomalies on the Higgs Branch of $\NN=2$ SCFTs}
\label{n2anomalies}

In addition to the $\half$-BPS Coulomb-branch operators, $\NN=2$ SCFTs possess another type of $\half$-BPS superconformal primary operators, which are neutral under the $U(1)_r$ but charged under the $SU(2)_R$ part of the R-symmetry group. These operators are called Higgs-branch operators (HBOs). The $\NN=2$ supersymmetric QFTs  have an associated moduli space of vacua---the Higgs branch---characterised by non-vanishing VEVs of these operators.

In \cite{Niarchos:2019onf} we focussed on $\NN=2$ SCFTs that have non-trivial superconformal manifolds and explored the properties of type-B Weyl anomalies for $\NN=2$ CBOs (as defined in Sec.~\ref{intro}) on the Higgs-branch moduli space. For definiteness, let us call $G^{\rm CFT}_{I\bar J}$ the type-B anomalies of CBOs $\OO_I$, $\bar \OO_J$ at the superconformal vacuum, and $G^{\IH}_{I\bar J}$ the corresponding value of these anomalies on the Higgs branch where conformal symmetry is spontaneously broken by default.

In \cite{Niarchos:2019onf,Andriolo:2022lcb} we argued that:
\begin{itemize}

\item[$(a)$] The anomalies $G^{\IH}_{I\bar J}$ are covariantly constant on the $\NN=2$ superconformal manifold
\beq
\label{eq:CovCostH}
\nabla_a G^{\IH}_{I\bar J} = 0
~.
\eeq
The index $a$ denotes collectively any coordinate on the superconformal manifold and $\nabla$ is the same connection as for the conformally symmetric phase. Since $\nabla$ is compatible with the (generalisation of the) Zamolodchikov metric, we also have by default that 
\beq
\label{eq:CovCostCFT}
\nabla_a G^{\rm CFT}_{I\bar J} =0 \, .
\eeq
These results are implied by requiring Wess--Zumino consistency for the corresponding anomaly functionals \cite{Andriolo:2022lcb}.

\item[$(b)$] The statement that both $G^{\rm CFT}_{I\bar J}$ and $G^{\IH}_{I\bar J}$ are covariantly constant on the superconformal manifold was used to make the argument that the anomalies match non-perturbatively in a finite region of the superconformal manifold if they match at one point. In other words, if there is a point $g^*$ on the superconformal manifold, where 
\beq
G^{\rm CFT}_{I\bar J}(g^*)=G^{\IH}_{I\bar J}(g^*)
\eeq can be established by independent computations, then it follows using item $(a)$ that the anomalies continue to match in a finite region of the superconformal manifold around $g^*$.

\end{itemize}

In \cite{Niarchos:2019onf} examples were presented where type-B anomalies for CBOs could be evaluated explicitly at weak coupling with a tree-level computation. In one of these examples (the 4D $\NN=2$ SCQCD theory) the anomalies for $\Delta = 2$ CBOs were found to match. In another example (the circular $\NN=2$ quiver) the corresponding anomalies for $\Delta = 2$ CBOs were initially reported in \cite{Niarchos:2019onf} to be different at and away from the origin of the Higgs branch, but a proper analysis shows that the anomalies match also in this case.

It is natural to ask:
\begin{itemize}

\item[$(\alpha)$] In which $\NN=2$ SCFTs and for which type-B CBO anomalies should one expect matching on the Higgs branch? 

\item[$(\beta)$] If type-B anomaly matching for CBOs across the Higgs branch can be established for the generators of the Coulomb-branch chiral ring, does it follow automatically that the corresponding anomalies match in the whole Coulomb-branch chiral ring?

\item[$(\gamma)$] A potential mismatch between $G^{\rm CFT}_{I\bar J}$ and $G^{\IH}_{I\bar J}$ is an interesting property. It implies that the $\NN=2$ superconformal manifold has a second rank-two symmetric tensor that is covariantly constant \eqref{eq:CovCostH},
besides the Hermitian metric $G^{\rm CFT}_{I\bar J}$ \eqref{eq:CovCostCFT}. Is this tensor a genuinely new two-tensor, or is it directly related to $G^{\rm CFT}_{I\bar J}$ (e.g.\ by a simple proportionality constant)? Can one extract scheme-independent quantities from these data that characterise the RG flow on the Higgs branch and what is their physical meaning?

\end{itemize}

In the next subsection we attempt to address the above questions with the formulation of two novel conjectures and summarise the preliminary evidence, presented in the rest of the paper, that supports them.

\subsection{Novel Conjectures for Type-B Anomaly Matching}
\label{conjectures}

Our first set of conjectures aims to address item $(\alpha)$ from the previous subsection. On the Higgs branch, part of the CFT spectrum becomes massive and therefore completely decouples in the extreme IR. This can include part (or all) of the CBOs. Let us call the set of CBOs in the UV theory that survive the RG flow the {\it IR chiral ring}.
We then make the following proposals depending on whether or not this set is empty: 

\paragraph{\bf Conjecture Ia.} {\it Trivial IR chiral ring: the type-B anomalies match along the Higgs-branch RG flow.}  

\vspace{0.5cm}
\noindent
For this class of theories all  type-B anomalies $G^{\IH}_{I\bar J}$ encode data of the massive spectrum. The conjectured matching relation $G^{\rm CFT}_{I\bar J}=G^{\IH}_{I\bar J}$  identifies the anomaly of the UV CFT phase with the corresponding anomaly along the RG flow. The $\NN=2$ SCQCD theory is an example that confirms this expectation. It will be discussed in detail in Sec.~\ref{sec:N2SCQCD}.

\paragraph{\bf Conjecture Ib.} {\it Non-trivial IR chiral ring: the type-B Weyl anomalies associated with CBOs in the IR chiral ring
match once the proper contribution of the massless IR degrees of freedom to the anomaly is taken into account. 
The anomalies for CBOs in the complement of the IR chiral ring are also expected to match. 
}

\vspace{0.5cm}
\noindent
When nontrivial IR chiral rings are present, the associated massless degrees of freedom can affect the matching of type-B anomalies. In Sec.~\ref{sec:quiver} we will consider the example of a circular $\NN=2$ superconformal quiver, which possesses a non-trivial IR chiral ring. We will show that the anomalies receive contributions from both massive and massless degrees of freedom along the RG flow. The anomalies of CBOs in the IR chiral ring match, as do the anomalies of CBOs in the complement,
in accordance with the above conjecture.

Next we move to item $(\beta)$ of Sec.~\ref{n2anomalies} regarding the role of the anomalies of the generators of the $\NN=2$ Coulomb-branch chiral ring. We propose:

\paragraph{\bf Conjecture II.} {\it A type-B anomaly matches if it involves CBOs generated (using the chiral ring OPE) by operators whose anomalies match.
}

\vspace{0.5cm}
\noindent
This conjecture is compatible with the previous Conjectures Ia and Ib. For simplicity, let us focus on $\NN=2$ SCFTs with a freely generated Coulomb-branch chiral ring.\footnote{In $\NN=2$ SCFTs the folklore is that the Coulomb-branch chiral ring is infinite dimensional, but freely generated \cite{Tachikawa2013n,Beem:2014zpa} by a finite number of CBOs. However, as outlined in \cite{Bourget:2018ond,Argyres:2018wxu,Bourton:2018jwb}, there exist some exceptions to this rule where the $\NN=2$ chiral ring is not generated freely. We do not expect that the existence of non-trivial chiral ring relations will affect the conclusion of our argument.} According to conjecture Ia, in the case of a trivial IR chiral ring all the type-B CBO anomalies are expected to match. This includes the anomalies of the generators. Conjecture II reasserts the matching of all anomalies as a consequence of the matching of the generators. In more general theories, the IR chiral ring is non-empty and conjecture II guarantees conjecture Ib, if one can establish the matching for the CBO anomalies of the ring generators. We will confirm Conjecture II by explicit computation in the example of $\NN=2$ SCQCD, which is discussed in Sec.~\ref{sec:N2SCQCD}.

\section{$\mathcal N=2$ SCQCD}\label{sec:N2SCQCD}

In the following two sections we proceed to show the above conjectures at work in explicit examples. Here we investigate the type-B anomalies of Coulomb-branch operators across the conformal and Higgs phases of 4D $\mathcal{N}=2$ superconformal QCD with gauge group $SU(K)$. This theory has a rich spectrum of CBOs with integer scaling dimension, which we will parametrise by 
\begin{equation}
    \Delta = 1 + L  \ \ \textrm{with} \ \ L \in \mathbb{Z}_{>0} \, \ .
\end{equation}
For convenience, we will focus on single-trace CBOs constructed from $\mathcal{N}=2$ vector multiplet scalars
\begin{align}
  \label{eq:30}
  \OO_{L+1} \propto \Tr[\varphi]^{L+1}\;.
\end{align}
Note, however, that the discussion can be trivially extended to multi-trace generalisations. The theory at the end of the RG flow initiated by the Higgs VEV has a trivial chiral ring, and the expectation from Conjecture Ia is that the anomalies in the conformal and Higgs phases should match.

As we have already discussed, on the one hand in the conformal phase of the theory the type-B anomaly arises as a logarithmic contribution to the two-point function in the momentum-space representation:
\begin{equation}
\label{eq:2point}
\langle \OO_{L+1} (p) \ \overbar \OO_{L+1} (-p) \rangle \, \ .
\end{equation}
On the other hand, in the Higgs phase of the theory  the corresponding type-B anomaly arises in a certain kinematical regime as a contribution to the three-point function,
\begin{equation}
\label{eq:3point}
\langle T(p) \ \OO_{L+1} (k_1) \ \overbar \OO_{L+1} (k_2) \rangle\;,
\end{equation}
where $T(p)$ denotes the trace of the energy-momentum tensor. As we will see shortly, even though the latter case involves complicated massive momentum integrals, the anomaly is contained within a simpler piece that can be evaluated analytically for all $L$.

\subsection{Computation of the Tree-level Anomaly in the CFT phase}
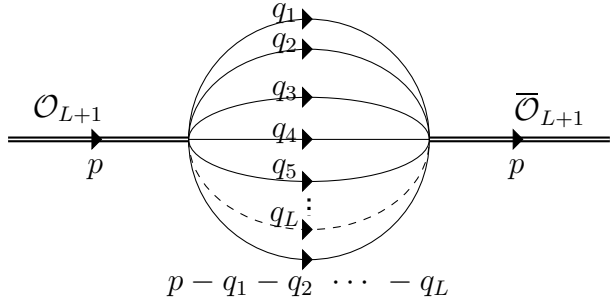
\begin{figure}
\center{
\begin{tikzpicture}[scale=0.8]

\draw[double,thick] (-2,0)-- node {\midarrow} (1,0);
\draw (1,0)-- node {\midarrow} (5,0);
\draw[double,thick] (5,0)-- node {\midarrow} (8,0);

\draw (3,1.95)-- node {\midarrow} (3,2.05);
\draw (3,-1.95)-- node {\midarrow} (3,-2.05);

\draw (3,1.45)-- node {\midarrow} (3,1.55);
\draw (3,-1.45)-- node {\midarrow} (3,-1.55);

\draw (3,0.65)-- node {\midarrow} (3,0.75);
\draw (3,-0.65)-- node {\midarrow} (3,-0.75);

\draw (3,0) circle (2cm);
\draw (3,0) ellipse (2cm and 0.7cm);
\draw (5,0) arc(0:180:2cm and 1.5cm);
\draw[dashed] (1,0) arc(180:360:2cm and 1.5cm);
\draw (-1,.5) node {$\OO_{L+1}$};
\draw (7,.5) node {$\overbar\OO_{L+1}$};

\node (a) at (-.55,-0.5) {$p$};
\node (b) at (6.45,-0.5) {$p$};

\node (c4) at (2.6,2.1) {$q_1$};
\node (c2) at (2.6,1.6) {$q_2$};
\node (c2) at (2.6,0.8) {$q_3$};
\node (c2) at (2.6,0.1) {$q_4$};
\node (c2) at (2.6,-0.5) {$q_5$};
\node (c2) at (2.6,-1.3) {$q_L$};

\node (cs) at (3,-0.8) {};
\node (cf) at (3,-1.45) {};

\draw[black, very thick, dotted] (cs) -- (cf);

\node (c3) at (3,-2.4) {$p-q_1-q_2 \  \cdots \ -q_{L}$};

\end{tikzpicture}
\caption{\it The diagram determining the leading contribution to the two-point function \eqref{eq:2point}.} \label{fig:massless}}
\end{figure}

We briefly review the computation of the anomaly $G_{L}^{CFT}$ in the CFT phase from the two-point function \eqref{eq:2point}. The two-point function at tree level corresponds to the Feynman diagram of Fig.~\ref{fig:massless}. It can be expressed as
\begin{align}
  \label{eq:9}
\langle \OO_{L+1} (p) \ \overline \OO_{L+1}(-p) \rangle  =   \mathcal{I}^{massless}_L(p) \CC_L^{CFT}\;,
\end{align}
where $\mathcal{I}^{massless}_L$ encodes the kinematical integral and  $\mathcal C_L^{CFT}$ the relevant colour factor
\begin{equation}
\label{eq:colorCFT}
    \mathcal{C}_L^{CFT} = \textrm{Tr}[T_{a_1} \cdots T_{a_{L+1}}]\sum_{\sigma \in S_{L+1}} \textrm{Tr}[T_{\sigma(a_1)}\cdots T_{\sigma(a_{L+1})}] \, \ ,
\end{equation}
where $\{T_{b}\}$, $b =1,\ldots,K^2 -1$, denote the generators of the Lie algebra $\mathfrak{su}(K)$. As all internal lines are simple scalar propagators, the kinematical contribution translates into the following momentum integral:\footnote{We use Euclidean signature in our Feynman diagram calculations throughout this paper.} 
\begin{equation}
\label{intCFT}
\mathcal{I}^{massless}_L(p) := \int \prod_{i=1}^{L}\frac{d^{4}q_i}{(2\pi)^4} \frac{1}{q_i^2} \times \frac{1}{(p-\sum_{i=1}^{L}q_i)^2} \, \ .
\end{equation}

We emphasise that, although the above expression involves $L$ momentum integrations, it is still capturing the leading (tree-level) contribution to the two-point function \eqref{eq:2point}. This integral suffers from UV divergences and needs to be regularised. In dimensional regularisation where $D = 4-2 \epsilon$ the answer is known, \emph{c.f.} \cite{Penati:2000zv,delaCruz:2019skx}, and reads
\begin{align}\label{eq:penati}
  \mathcal{I}^{massless}_L(p) =\frac{(-1)^{L+1}}{(L!)^2}\frac{(p^2)^{(L-1)}}{(4 \pi)^{2L}} \Big(  \frac{1}{\epsilon}  - L \log{p^2} + O(\epsilon^0)\Big) \;.
\end{align}
The anomaly is extracted from this result by isolating the logarithmic contribution according to Eq.~\eqref{introaa}. In this fashion one arrives at the  simple result
\begin{align}
  \label{eq:10}
G_L^{CFT}  = \frac{\CC_L^{CFT}}{(2\pi)^{2L+2}}\;.
\end{align}

\subsection{Alternative Expression for the Massless Integral}\label{sec:intermediate}

Although we have completed the anomaly calculation in the conformal phase, it will be useful to highlight here an intermediate mathematical result by rewriting \eqref{intCFT} using Feynman parametrisation. This result will play a role in the subsequent evaluation of the anomaly in the broken phase. To that end, let us briefly set out a few relevant conventions.

We will write products of propagators using the well-known identity
\begin{equation}
\label{id1}
 \frac{1}{D_1D_2...D_N} = \int_{0}^{1}dx_1...\int_{0}^{1}dx_N \ \delta\left(\sum_{i=1}^N x_i -1\right)\frac{(N-1)!}{[x_1D_1+x_2D_2+...x_ND_N]^N}\;,
\end{equation}
where $\{x_i\}$ are Feynman parameters. The integration over the set of internal momenta $\{q_i\}$ can be performed in a recursive way, by completing the corresponding square and applying the identity 
\begin{equation}
\label{eq:int}
\int \frac{d^Dl}{(2\pi)^D}\frac{1}{(l^2-\Delta)^N} = \frac{(-1)^N i}{(4\pi)^{D/2}}\frac{\Gamma(N-D/2)}{\Gamma(N)}\left(\frac{1}{\Delta}\right)^{N-D/2}\;.
\end{equation}
Using the above relations, one can recast  $L$-loop integrals into the general form
\begin{equation}
\label{eq:J}
\mathcal{J} = (N-1)!\int_0^1 \prod_{j=1}^{N}dx_j\delta\left(\sum_{i=1}^{N}x_i-1\right)\int \prod_{i=1}^{L} \frac{d^Dq_i}{(2\pi)^D}\left[\sum_{i=1}^{L}q_iq_jM_{ij} -2 \sum_{j=1}^{L}q_jK_j +J \right]^{-N} \, \ ,
\end{equation}
where $N$ is the number of Feynman parameters. The integrations over all internal momenta can be carried out employing  (\ref{eq:int}) to obtain \cite{Weinzierl:2006qs}
\begin{equation}
\label{eq:result}
\mathcal{J} =\frac{ (-1)^{N}\Gamma(N-LD/2)}{(4\pi)^{DL/2}}\int_0^1 \prod_{j=1}^{N}dx_j \ \delta\left(\sum_{i=1}^N x_i -1\right)\frac{\mathcal{U}^{N-(L+1)D/2}}{\mathcal{F}^{N-LD/2}} \, \ ,
\end{equation}
where
\begin{equation}
\label{eq:uf}
 \mathcal{U} := \textrm{Det}[M]\;, \qquad \mathcal{F} := \textrm{Det}[M]\left( \sum_{i,j=1}^{L}K_iM^{-1}_{ij}K_j  - J\right)   \;.
\end{equation}

We can apply this parametrisation directly to the massless integral \eqref{intCFT}, where for now we keep the spacetime dimension $D$ generic in light of the fact that we will be using dimensional regularisation. Through formula \eqref{eq:J} the corresponding quantity $M^{massless}$ is determined to be the $L \times L$ matrix 
\begin{align}
& M^{massless} = \begin{pmatrix}
     x_1+x_{L+1} & x_{L+1} & x_{L+1} & \cdots & x_{L+1} \\
     x_{L+1} & x_2+ x_{L+1} & x_{L+1} & \cdots & x_{L+1} \\
     x_{L+1} & x_{L+1} & x_3+ x_{L+1} & \cdots & x_{L+1} \\
     \vdots & \vdots & \vdots & \ddots & \vdots \\
     x_{L+1} & x_{L+1} & x_{L+1}  & \cdots & x_{L} + x_{L+1}
     \end{pmatrix} \, \ .
\end{align}
Using (\ref{eq:uf}) one also computes all the other quantities
\begin{align}\label{Umassless}
& K^{massless} = (\underbrace{px_{L+1}, \cdots ,px_{L+1}}_{L \ \textrm{times}}), \ \ \ \ J^{massless} = p^2x_{L+1} \, ,\cr
& \mathcal{U}^{massless}(x_j)  = \sum_{1\leq i_1 < i_2 < \cdots <i_{L-1} <i_{L} \leq L+1}x_{i_1}x_{i_2} \cdots x_{i_{L-1}}x_{i_{L}}  \, \ ,  \cr
& \mathcal{F}^{massless}(p^2,x_j) = -p^2\prod_{i=1}^{L+1}x_i     \, \ ,
\end{align} 
in terms of which \eqref{eq:result} becomes
\begin{align}
   \mathcal{I}^{massless}_{L}(p) =& \frac{(-1)^{LD/2}\Gamma(L(1-D/2)+1)}{(4\pi)^{DL/2}}(p^2)^{L(D/2-1)-1} \times \\ \nonumber
&\qquad \times\int_0^1 \prod_{j=1}^{L+1}dx_j \ \delta\left(\sum_{i=1}^{L+1} x_i -1\right)  \frac{\mathcal{U}^{massless}(x_k)^{(L+1)(1-D/2)}}{(\prod_{i=1}^{L+1}x_i)^{L(1-D/2)+1}} \, \ .
\end{align}
We then set $D=4-2\epsilon$ and perform a series expansion in $\epsilon$. Let us focus on the logarithmic term
\begin{equation}
\mathcal{I}_{L}^{massless}\simeq  c^{massless}_L \log{p^2}    \, \ ,
\end{equation}
where
\begin{align}
   c^{massless}_L:=\frac{(-1)^{L}}{(L-1)!}\frac{(p^2)^{L-1}}{(4\pi)^{2L}}\int_0^1 \prod_{j=1}^{L+1}dx_j \ \delta\left(\sum_{i=1}^{L+1} x_i -1\right) \frac{\left(\prod_{i=1}^{L+1}x_i\right)^{L-1}}{\mathcal{U}^{massless}(x_k)^{L+1}} \, \ .
\end{align}
One can easily perform the integration over the variable $x_1$ to obtain
\begin{align}
  c_{L}^{massless} = \frac{(-1)^{L}}{(L-1)!}\frac{(p^2)^{L-1}}{(4\pi)^{2L}}&\int_0^1 dx_2 \int_0^{1-x_2} dx_3  \cdots\cr
  &\cdots \int_0^{1-x_2 \cdots -x_{L}} dx_{L+1} \  \frac{\left(\left(1-\sum_{i=2}^{L+1}x_i\right)\prod_{i=2}^{L+1}x_i\right)^{L-1}}{f(x_j)^{L+1}} \, ,
\end{align}
with
\begin{equation}
\label{eq:f}
    f(x_k) := \mathcal{U}^{massless}(x_k)|_{x_1=1-\sum_{i=2}^{{L+1}}x_i} \ \ \textrm{for} \ \ k=2, \ldots, L+1 \, \ .
\end{equation}
As a last step we choose to shift $x_{j+1} \mapsto x_j$ for $j=1,\ldots,L$ to arrive at 
 \begin{align}
    \label{eq:masslessfinal}
    c_{L}^{massless} = \frac{(-1)^{L}}{(L-1)!}\frac{(p^2)^{L-1}}{(4\pi)^{2L}}\int_0^1 dx_1 \int_0^{1-x_1} dx_2  \cdots \int_0^{1-x_1 \cdots -x_{L-1}} dx_L \   B(x_1, \ldots, x_{L})\;,
\end{align} 
where 
\begin{align}\label{eq:integrand}
B(x_1, \ldots, x_{L}):=  \frac{\left(\left(1-\sum_{i=1}^{L}x_i\right)\prod_{i=1}^{L}x_i\right)^{L-1}}{f(x_j)^{L+1}}\;.
\end{align}
Note that this is a symmetric function under an exchange of the parameters $x_i$.

We can finally compare \eqref{eq:masslessfinal} with the $\log{p^2}$ coefficient of  \eqref{eq:penati} to deduce that
\begin{align}\label{eq:masslessexplicit}
  \int_0^1 dx_1 \int_0^{1-x_1} dx_2  \cdots \int_0^{1-x_1 \cdots -x_{L-1}} dx_L \  B(x_1, \ldots, x_{L}) = \frac{1}{L!}\;.
\end{align}
It is this result that will be used in the upcoming evaluation of the anomaly in the broken phase.

\subsection{Computation of  the Tree-level Anomaly in the Higgs phase}\label{SCQCDHiggs}

\begin{figure}
\center{
\begin{tikzpicture}[scale=0.8]

\draw[double,thick,snake it] (-2,0) to (0,0);
\draw[dashed] (0,0) -- node {\midarrow} (2,0);
\draw (2,0) -- node {\miduno} (4.5,2.6) ;
\draw[double,thick] (4.5,2.6) -- node {\miduno}(6,4.1);
\draw (2,0) -- node{\middue}(4.5,-2.6);
\draw[double,thick] (4.5,-2.6) -- node{\middue}(6,-4.1);
\draw (-1.5,0.5) node {$T$};
\draw (-1.5,-.5) node {$p$};
\filldraw (0,0) circle (2pt);
\draw (1,0.5) node {$\sigma$};

\node (a)  at (4.5,2.6) {};
\node (b)  at (4.5,-2.6) {};

\node (c) at (6,0) {};
\node (c1) at (6,-0.3) {};

\node (d) at (6.5,0) {};
\node (d1) at (6.5,-0.3) {};

\draw (4.5,0) -- (a);
\draw (b) -- node {\midup} (4.5,0) ;

\path (b) [bend left] edge node [above] {\midup}  (a);
\path (b) [bend right] edge node [above] {\midup}  (a);

\path[dashed] (a) [bend left] edge node [above] {} (c1);
\path[dashed] (b) [bend right] edge node [above] {\middiag} (c);

\path (a) [bend left] edge node [above] {} (d1);
\path (b) [bend right] edge node [above] {} (d);
\path (6.5,-0.1) -- node {\midup} (d);

\node (c4) at (2.9,0) {$q_2-q_1$};
\node (c4) at (4.6,-1) {$q_3-q_2$};
\node (c4) at (5.3,0.5) {$q_4-q_3$};
\node (c4) at (7,-1.7) {$q_L-q_{L-1}$};
\node (c4) at (7.4,-.23) {$k-q_{L}$};

\node (c1) at (1,-.5) {$p$};

\node (c1) at (3,1.8) {$q_1$};
\node (c1) at (3,-1.8) {$p-q_1$};

\node (c1) at (6,4.5) {$k_1:=k$};
\node (c1) at (6,-4.5) {$k_2 = p-k_1$};

\node (ff) at (5.6,-0.5) {$.....$};

\draw (6.55,-3.5) node {$\overbar \OO_{L+1}$};
\draw (6.55,3.5) node {$\OO_{L+1}$};

\end{tikzpicture}
\caption{\it The diagram determining the leading contribution to the three-point function \eqref{eq:3point}.}
\label{fig:massive}
}
\end{figure}
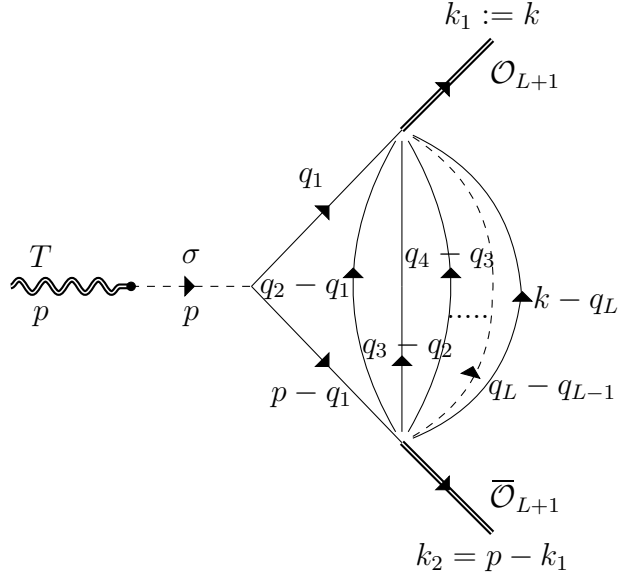

We now turn to the computation of the anomaly in the Higgs phase, where one of the fundamental hypermultiplet scalars acquires a VEV proportional to the parameter $v$. This renders the adjoint scalars massive with $m^2 = 2 v^2 g^2$ in the conventions of  \cite{Niarchos:2019onf}. The three-point function \eqref{eq:3point} is captured at tree level by the Feynman diagram of Fig.~\ref{fig:massive}. We use the rules given in Sec.~7.1 of \cite{Niarchos:2019onf} to evaluate this diagram, and point the interested reader to that reference for a detailed discussion on how they are derived.

In summary:
\begin{itemize}
\item The linear coupling between the trace of the energy-momentum tensor and the dilaton contributes a factor of $v p^2/2$.
\item The dilaton propagator gives a factor of $2iK/p^2$. 
\item There is a factor of $(-i)(-\frac{2}{K}g^2 v)$ from the vertex $\sigma \varphi \bar\varphi$.
\item There is an $L$-loop-momentum integral involving the scalar propagators
\begin{align}
\label{intHIGGS}
\mathcal{I}^{massive}_L(k,p) := &\int \prod_{j=1}^{L} \frac{d^4q_j}{(2\pi)^4} \frac{1}{q_1^2-m^2}\frac{1}{(p-q_1)^2-m^2}\times\cr
&\qquad\qquad\times\prod_{i=1}^{L-1}\frac{1}{(q_{i+1}-q_{i})^2-m^2}\frac{1}{(k-q_L)^2-m^2}  \;,
\end{align}
where we have used the conservation of external momenta and have defined $k$ as
\begin{equation}
    p = k_1+ k_2  \ \ \Rightarrow \ \  k_2 = p-k_1 = p-k \, \ .
  \end{equation}
\item There is a colour factor $\CC_L^{\mathbb H}$. In conventions where
  $\Tr[T_a T_b ] = \delta_{ab}$ with $a,b = 1,\ldots, K^2 -1$, we have:
\begin{align}\label{colourfactors}
  \mathcal{C}_L^{\mathbb H} & = \sum_{\pi \in \mathbb Z_{L+1}}\textrm{Tr}[T_{\pi(a_1)}T_{b}]\textrm{Tr}[T_{b}T_{\pi(a_2)} \cdots T_{\pi(a_{L+1})}]\sum_{\sigma \in S_{L+1}} \textrm{Tr}[T_{\sigma(a_1)}T_{\sigma(a_2)}\cdots T_{\sigma(a_{L+1})}]  \cr
  &= (L+1)\textrm{Tr}[T_{a_1}T_{b}]\textrm{Tr}[T_{b}T_{a_2} \cdots T_{a_{L+1}}]\sum_{\sigma \in S_{L+1}} \textrm{Tr}[T_{\sigma(a_1)}T_{\sigma(a_2)}\cdots T_{\sigma(a_{L+1})}]  \cr
    & = (L+1) \CC_L^{CFT}\; .
\end{align}

\end{itemize}

Putting everything together, the three-point function of interest can be expressed as
\begin{align}
  \label{eq:12}
      \langle T(p) \ \OO_{L+1} (k_1) \ \overline \OO_{L+1} (k_2) \rangle = - (L+1) m^2 \mathcal{I}^{massive}_L \CC_L^{CFT}\;.
\end{align}
The prescription for extracting the anomaly from the three-point function has been given in \eqref{introba} and reads
\begin{align}
  \label{eq:13}
  G_{L}^{\mathbb H}=& (-1)^{L+1}\frac{2^{2L-2}\Gamma(L)\Gamma(L+1)}{\pi^2((L-1)!)^2}\times\cr
              &\qquad \qquad\lim_{p\to 0}\lim_{k_1, k_2\to 0}\Big[\frac{d^{L-1}}{dk_1^{L-1}}\frac{d^{L-1}}{dk_2^{L-1}}    \langle T(p) \ \OO_{L+1}  (k_1) \ \overline \OO_{L+1} (k_2) \rangle\Big]\;.
\end{align}
Note, however, that when we impose the conservation of the external momenta (so that $\vec k_2 = \vec p -\vec k_1$) in the limit of the momentum magnitudes $p\to 0$ and $k\to 0$---where $k_2 \to k_1 =: k $---we can combine \eqref{eq:12} and \eqref{eq:13} to get
\begin{align}
  \label{eq:14}
  G_{L}^{\mathbb H}= (L+1) m^2  \CC_L^{CFT}  (-1)^{L}\frac{2^{2L-2}\Gamma(L)\Gamma(L+1)}{\pi^2(2L-2)!}\lim_{p\to 0}\lim_{k\to 0}\Big[\frac{d^{2L-2}}{dk^{2L-2}}  \mathcal{I}^{massive}_L\Big]\;.
\end{align}

Our next task in determining \eqref{eq:14} is the evaluation of $\mathcal{I}^{massive}_L$. To the best of our knowledge, ready to use analytic expressions for this integral are not known for general $L$, even though such integrals are the subject of a very active area of research \cite{Bogner:2007mn,Bourjaily:2018ycu,Bourjaily:2018yfy,Broedel:2019kmn,Bourjaily:2019hmc,Klausen:2019hrg,Klemm:2019dbm, Hidding:2020ytt,Bonisch:2020qmm}. We will now see that  the anomaly is associated with a relatively-simple piece of the integral that is calculable in the requisite momentum limits.

We find the use of the Feynman parametrisation that we introduced in Sec.~\ref{sec:intermediate} essential for this task. To proceed, we recast the original version of $\mathcal{I}^{massive}_L$ from (\ref{intHIGGS}) in the form \eqref{eq:J}. From the latter one can read off
\begin{align}\label{Mmassive}
    M^{massive} = \begin{pmatrix}
    x_1 +x_2 +x_3 & -x_3 & 0 & 0 & \cdots & 0 \\
    -x_3 & x_3 +x_4 & -x_4 & 0 & \cdots & 0 \\
    0 & -x_4 & x_4+x_5 & -x_5  & \cdots & 0 \\
    \vdots & \vdots & \vdots & \ddots & \cdots & 0\\
    0 &  \cdots & 0 & -x_{L} & x_{L}+x_{L+1} & -x_{L+1} \\
    0 & 0 & \cdots & 0 & -x_{L+1} & x_{L+1}+x_{L+2}
\end{pmatrix} \, \ .
\end{align}
Using (\ref{eq:uf}) we then compute the remaining quantities 
\begin{align}\label{UKJmassive}
  \mathcal{U}^{massive}(x_j) =& \sum_{1\leq i_1 < \cdots <i_{L} \leq L+2}x_{i_1}x_{i_2} \cdots x_{i_{L-1}}x_{i_{L}} -  x_1x_2\left(\sum_{3\leq j_1  < \cdots <j_{L-2} \leq L+2}x_{j_1}x_{j_2} \cdots x_{j_{L-3}}x_{j_{L-2}}\right) \,\cr
                                K^{massive}=&(px_2, \underbrace{0, \cdots, 0,}_{L-2 \ \textrm{times}} kx_{L+2} ), \qquad J^{massive} = -m^2 +p^2x_2+k^2x_{L+2} 
\end{align}
and
    \begin{align}   \label{UKJFdefsmassive}
 \mathcal{F}^{massive}(k,p,m^2,x_j) = & m^2f(x_{j+1}) -p^2\left(x_1x_2\sum_{3\leq j_3 < \cdots < j_{L+1}\leq L+2}x_{j_3}x_{j_4}\cdots x_{j_{L}}x_{j_{L+1}}+ \prod_{i=2}^{L+2}x_j\right)   \cr
&\qquad+ 2pk\prod_{i=2}^{L+2}x_i -k^2(x_1+x_2)\prod_{j=3}^{L+2}x_j \, , 
\end{align}
where the function $f(x_j)$ was defined through \eqref{eq:f} and as a result $f(x_{j+1})$ has $j=3,\ldots,L+2$. In analogy with the massless-case discussion, we can use \eqref{eq:result} to express $\mathcal{I}^{massive}_L(k,p)$ as
\begin{equation}
\mathcal{I}^{massive}_L =\frac{ (-1)^{L+2}\Gamma(L(1-D/2)+2)}{(4\pi)^{DL/2}}\int_0^1 \prod_{j=1}^{L+2}dx_j \ \delta\left(\sum_{i=1}^{L+2} x_i -1\right)\frac{(\mathcal{U}^{massive})^{L(1-D/2)+(2-D/2)}}{(\mathcal{F}^{massive})^{L(1-D/2)+2}} \, \ .
\end{equation}

It is again necessary to implement dimensional regularisation by setting $D=4-2\epsilon$ and performing a series expansion in $\epsilon$, resulting in
\begin{align}
\mathcal{I}^{massive}_L = \frac{1}{\epsilon} \; c^L_{-1} + c^L_0 +\epsilon \;c^L_1 + O(\epsilon) \, \ .
\end{align}
We note that the $\frac{1}{\epsilon}$ coefficient is given by
\begin{align}
c^L_{-1} \propto \frac {(\mathcal{F}^{massive})^{L-2}}{(\mathcal{U}^{massive})^{L}}\quad\textrm{for}\quad L\ge 2\;,
\end{align}
where by \eqref{UKJFdefsmassive} the above expression is a polynomial of degree $2L-4$ in $k$.\footnote{For $L=1$ one has that $c_{-1}^1 = 0$.} 

The anomaly is obtained by taking $(2L-2)$ $k$ derivatives of $\II_L^{massive}$ as in equation \eqref{eq:14}.
The derivatives kill the $c^L_{-1}$ contribution and
thus we can safely take the $\epsilon\to 0$ limit obtaining a contribution only from the $c_0^L$ term. As a result
\begin{align}
  & \lim_{p \rightarrow 0} \ \lim_{k \rightarrow 0}\left[\frac{d^{2L-2} \ \II_L^{massive}(k,p)}{d k^{2L-2}}  \right] = \lim_{p \rightarrow 0} \ \lim_{k \rightarrow 0}\left[\frac{d^{2L-2} \ c_0^L(k,p)}{d k^{2L-2}} \right] \cr
&\qquad\quad= \int_0^1 \prod_{j=1}^{L+2} dx_j \ \delta\left(1-\sum_{i=1}^{L+2}x_j\right) \frac{(-1)^L(2L-2)!}{(4\pi)^{2L}(L-1)!m^2}\frac{\left[(x_1+x_{2})\prod_{j=3}^{L+2}x_j\right]^{L-1}}{\mathcal{U}^{massive}(x_j)^{L+1}} \; .
\end{align}
This integral can now be evaluated as follows. We choose to integrate over $x_{2}$ using the $\delta$-function, and rename the integration variables $x_{j+2} \mapsto x_j$ for $j=1,\ldots,L$ and $x_1\mapsto u$. In this way the above expression can be massaged into 
\begin{align}\label{eq:massivef}
  \lim_{p \rightarrow 0} \ \lim_{k \rightarrow 0}\left[\frac{d^{2L-2}\ \II_L^{massive}(k,p)}{d k^{2L-2}}\right] = & \frac{(-1)^L(2L-2)!}{(4\pi)^{2L}(L-1)!m^2} \times\cr
   & \times \int_0^1 du \int_0^{1-u} dx_1 \cdots \int_0^{1-u-x_1 \cdots -x_{L-1}} dx_L  B(x_1, \ldots , x_L)\, \ ,\cr
\end{align}
where $B(x_1, \ldots , x_L)$ was defined in \eqref{eq:integrand}. One can use the Fubini--Tonelli theorem to iteratively move the $u$ integration behind the $x_i$ integrations such that
\begin{align}\label{starting}
&  \int_0^1 du \int_0^{1-u} dx_1  \cdots  \int_0^{1-u-x_1 \cdots -x_{L-1}} dx_L  B(x_1, \ldots , x_L) \cr
  & =   \int_0^1 dx_1 \int_0^{1-x_1} dx_2 \cdots \int_0^{1-x_1 \cdots -x_{L-1}} dx_L  (1-x_1-\cdots - x_L)B(x_1, \ldots , x_L)   \;.
\end{align}
We can then introduce a redundant $\delta$-function to  re-write the integral as
\begin{align}\label{LHS}
  & \int_0^1 \prod_{i=1}^{L+1}dx_i  \delta(1-x_1 - \cdots -x_{L+1})  (1-x_1-\cdots - x_L)B(x_1, \ldots , x_L) \cr
  & =\int_0^1 \prod_{i=1}^{L+1}dx_i  \delta(1-x_1 - \cdots -x_{L+1})  (1-L x_1) \frac{(\prod_{i = 1}^{L+1}x_i)^{L-1}}{\mathcal U^{massless}(x_j)^{L+1}}\cr
  & = \frac{1}{L!} - L \int_0^1 \prod_{i=1}^{L+1}dx_i  \delta(1-x_1 - \cdots -x_{L+1}) x_1 \frac{(\prod_{i = 1}^{L+1}x_i)^{L-1}}{\mathcal U^{massless}(x_j)^{L+1}} \;,
\end{align}
where in the second line we expressed the function $B(x_1, \ldots, x_L)$ in terms of $\mathcal U^{massless}(x_j)$ as defined  in \eqref{Umassless}. We also made use of the fact that this fraction is symmetric under the exchange of all  the $x_i$. In the last line we recognised the  massless integral \eqref{eq:masslessexplicit}, which returns the value $\frac{1}{L!}$.

Note that, using the same starting point as in \eqref{LHS}, we can also write
\begin{align}\label{RHS}
  \int_0^1 \prod_{i=1}^{L+1}&dx_i  \delta(1-x_1 - \cdots -x_{L+1})  (1-x_1-\cdots - x_L)B(x_1, \ldots , x_L) \cr
  & =  \int_0^1 \prod_{i=1}^{L+1}dx_i  \delta(1-x_1 - \cdots -x_{L+1}) x_{L+1}\frac{(\prod_{i = 1}^{L+1}x_i)^{L-1}}{\mathcal U^{massless}(x_j)^{L+1}}\cr
       & =  \int_0^1 \prod_{i=1}^{L+1}dx_i  \delta(1-x_1 - \cdots -x_{L+1}) x_{1}\frac{(\prod_{i = 1}^{L+1}x_i)^{L-1}}{\mathcal U^{massless}(x_j)^{L+1}}\;,
\end{align}
where in the last line we once again used the symmetry property of the fraction under permutations of all the $x_i$.

Finally, equating \eqref{LHS} and \eqref{RHS} yields
\begin{align}
 \int_0^1 \prod_{i=1}^{L+1}dx_i  \delta(1-x_1 - \cdots -x_{L+1}) x_{1}\frac{(\prod_{i = 1}^{L+1}x_i)^{L-1}}{\mathcal U^{massless}(x_j)^{L+1}} = \frac{1}{(L+1)!}\;,
\end{align}
which can be used to obtain a simple expression for \eqref{eq:massivef}
\begin{align}\label{cmass}
 \lim_{p \rightarrow 0} \ \lim_{k \rightarrow 0}\left[\frac{d^{2L-2} \ \II_L^{massive}(k,p)}{d k^{2L-2}} \right] =  \frac{(-1)^L(2L-2)!}{(4\pi)^{2L}(L-1)!(L+1)!m^2} \, \ .  
\end{align}

It is now straightforward to combine all the factors in \eqref{eq:14} to arrive at
\begin{align}
  \label{eq:15}
  G_{L}^{\mathbb H}&= m^2  \CC_L^H  (-1)^{L}\frac{2^{2L-2}\Gamma(L)\Gamma(L+1)}{\pi^2(2L-2)!}  \lim_{p \rightarrow 0} \ \lim_{k \rightarrow 0}\left[\frac{d^{2L-2} \ \II_L^{massive}(k,p)}{d k^{2L-2}} \right]\cr
  & = \frac{\CC_L^{CFT}}{(2\pi)^{2L+2}} \;.
\end{align}
This expression agrees precisely with the tree-level anomaly in the CFT phase of the theory as evaluated in \eqref{eq:10}. Combined with the fact that $\nabla G_L^{CFT} = \nabla G_L^{\mathbb H} = 0$, one determines that the anomalies match non-perturbatively for finite values of the exactly marginal coupling. This result generalises that of \cite{Niarchos:2019onf} from $L=1$ to all values of $L$. We emphasise that, even though in the above example we considered single-trace CBOs, the anomaly matching extends to the case of multi-trace CBOs, essentially because the colour structure always appears implicitly through \eqref{colourfactors}.

\subsection{The $L=2$ Massive Integral as a Sunset Integral}

As a supplement to the above discussion, in this section we present an alternative approach to the evaluation of the anomaly in the broken phase, without the use of Feynman parametrisation for the kinematical factor. For simplicity we focus on the specific case of $L=2$ and the $\Tr[\varphi^3]$ CBO, which leads to $\II_{2}^{massive}$; this is an example of the well studied `sunset integral'. For $L=3$ one could in principle repeat this procedure using the three-loop equal-mass `banana-amplitude' results of \cite{Klemm:2019dbm,Hidding:2020ytt}, although that would be a technically hard exercise which would not provide novel insights for our purposes. For $L>3$, the situation is even more challenging: although higher-loop banana integrals can be written in terms of periods of families of Barth--Nieto Calabi--Yau manifolds \cite{Klemm:2019dbm}, we are not aware of ready-to-use results which could be employed for the calculation of our massive integral.

We would therefore like to re-evaluate the kinematical piece for $L=2$ in the anomaly \eqref{eq:13}, that is
\begin{align}
G_2^{\mathbb H} \propto \lim_{p\to 0}\lim_{k\to 0}\Big[\frac{d^{2}}{dk^{2}}  \mathcal{I}^{massive}_2\Big]\;,
\end{align}
where after a  redefinition $q_i\mapsto \sum_{j=1}^i q_j$, and in the $p\to 0$ limit, the massive integral of Eq.~\eqref{intHIGGS} reads for $L=2$
\begin{align}
\label{intHIGGSL2}
\lim_{p\to 0}\mathcal{I}^{massive}_2(k,p) = &\int \prod_{j=1}^{2} \frac{d^4q_j}{(2\pi)^4} \frac{1}{(q_1^2-m^2)^2}\frac{1}{q_{2}^2-m^2}\frac{1}{(k-q_1 -q_2)^2-m^2}  \;.
\end{align}
This is a particular example of a `sunset integral', defined as 
\begin{align}
\label{sunset}
I_{\alpha_1 \alpha_2 \alpha_3}(k^2,m_1^2,m_2^2,m_3^2) := &\int \frac{d^4q_1}{(2\pi)^4}\frac{d^4q_2}{(2\pi)^4} \frac{1}{(q_1^2-m^2)^{\alpha_1} }\frac{1}{(q_{2}^2-m^2)^{\alpha_2}}\frac{1}{((k-q_1 -q_2)^2-m^2)^{\alpha_3}}  \;,
\end{align}
for $\alpha_1 = 2,~ \alpha_2 = \alpha_3 = 1$ and $ m_1^2 = m_2^2 =m_3^2= m^2$. This integral can be determined from the simpler $I_{111}$ through the relation \cite{Laporta:2004rb,Tarasov:2006nk}
\begin{align}
  \label{I211}
  I_{211}(k^2,m^2) = \frac{1}{3}\frac{\partial}{\partial m^2} I_{111}(k^2,m^2)\;.
\end{align}
One can compute the UV-divergent $I_{111}$ in dimensional regularisation, obtaining  \cite{Bloch:2013tra}
\begin{align}
  \label{eq:i111}
  I_{111}(k^2,m^2) = 16 \pi^{4-2\epsilon}\Gamma(1+\epsilon)^2\left(m^2\right)^{1-2\epsilon}\left(\frac{a_2}{\epsilon^2}+\frac{a_1}{\epsilon}+a_0+ O(\epsilon)\right) \, ,
\end{align}
with the coefficients in the $\epsilon$ expansion given by
\begin{align}
  & a_2 = -\frac{3}{8}, \cr
  & a_1=\frac{18-t}{32}, \cr
  & a_0 = \frac{(t-1)(t-9)}{12}\left(1+(t+3)\frac{d}{dt}\right)\mathcal{J}^2(t) + \frac{13t-72}{128}\;.
\end{align}
In the above expressions $t = \frac{k^2}{m^2}$ and $\mathcal{J}^2(t)$ (for $t < 9$) is given by \cite{Bloch:2013tra}
\begin{equation}
\mathcal{J}^2(t) = \sum_{n \geq 0}J_nt^n  \; , \ \ \  J_n =\frac{1}{4^{n-1}n!^2}\int_{0}^{\infty}x^{2n+1}K_{0}(x)^3 dx\;,
\end{equation}
with $K_0(x)$ the modified Bessel function of the second kind.

Eq.~\eqref{I211} can then be directly evaluated, yielding
\begin{align}
  \label{eq:18}
  \frac{\partial }{\partial m^2}I_{111} =&~\frac{1}{(2\pi)^4}\Big\{ \frac{1}{\epsilon}\Big[-\frac{3}{2}a_2- 2 a_2 (1 + \gamma + \log{(\pi m^2)})\Big]\cr
  & +\frac{\partial }{\partial m^2}\Big[\frac{m^2}{6} [6 a_0  + 12 \gamma(-a_1 + a_2 \gamma) + a_2 \pi^2 \cr
  &\quad +  12\log{(\pi m^2)}(-a_1  + 2 a_2 \gamma  + a_2 \log{(\pi m^2)})] \Big]\Big\}  + O(\epsilon^2)\;.
\end{align}
We use this explicit expression to extract the following simple result in the $\epsilon\to 0 $ limit
\begin{align}
  \label{eq:1}
  \lim_{k\to 0 } \frac{\partial^2}{\partial k^2}I_{211} = \frac{1}{(2 \pi)^4}\frac{1}{48 m^2}\;,
\end{align}
which does not contribute a divergent piece to the anomaly since the $O(\frac{1}{\epsilon})$ term in \eqref{eq:18} is independent of $t$, and therefore $k^2$. This expression agrees exactly with the $L=2$ result from \eqref{cmass}.

\section{$\NN=2 $ Circular Quiver}\label{sec:quiver}

\begin{figure}
\begin{center}
\begin{tikzpicture}[scale=0.7]

\def\circledarrow#1#2#3{ 
\draw[#1,->] (#2) +(80:#3) arc(80:-260:#3);
}
 

\draw (0,3) circle [radius=0.5]  node (A) {$K$};
\draw (1.5*1.414,1.5*1.414)  circle [radius=0.5] node {$K$};
\draw (3,0)  circle [radius=0.5] node {$K$};
\draw (1.5*1.414,-1.5*1.414)  circle [radius=0.5] node {$K$};
\draw (0,-3)  circle [radius=0.5] node {$K$};
\draw (-1.5*1.414,-1.5*1.414)  circle [radius=0.5] node {$K$};
\draw (-3,0)  circle [radius=0.5] node {$K$};
\draw (-1.5*1.414,1.5*1.414)  circle [radius=0.5] node {$K$};


\draw [->,thick,domain=54:81,scale=3] plot ({1.05*cos(\x)}, {1.05*sin(\x)}) node[above right] {$\quad\widetilde{Q}^{(1)}$};
\draw [->,thick,domain=81:54,scale=3] plot ({0.95*cos(\x)}, {0.95*sin(\x)}) node[below left] {$Q^{(1)}$};;

\draw [->,thick,domain=9:36,scale=3] plot ({1.05*cos(\x)}, {1.05*sin(\x)});
\draw [->,thick,domain=35:9,scale=3] plot ({0.95*cos(\x)}, {0.95*sin(\x)});

\draw [<-,thick,domain=-9:-36,scale=3] plot ({1.05*cos(\x)}, {1.05*sin(\x)});
\draw [<-,thick,domain=-36:-9,scale=3] plot ({0.95*cos(\x)}, {0.95*sin(\x)});

\draw [<-,thick,domain=-54:-80,scale=3] plot ({1.05*cos(\x)}, {1.05*sin(\x)});
\draw [<-,thick,domain=-80:-54,scale=3] plot ({0.95*cos(\x)}, {0.95*sin(\x)});

\draw [->,thick,domain=54:80,scale=-3] plot ({1.05*cos(\x)}, {1.05*sin(\x)});
\draw [->,thick,domain=80:54,scale=-3] plot ({0.95*cos(\x)}, {0.95*sin(\x)});

\draw [->,thick,domain=9:35,scale=-3] plot ({1.05*cos(\x)}, {1.05*sin(\x)});
\draw [->,thick,domain=35:9,scale=-3] plot ({0.95*cos(\x)}, {0.95*sin(\x)});

\draw [->,dashed,thick,domain=-35:-9,scale=-3] plot ({1.05*cos(\x)}, {1.05*sin(\x)});
\draw [->,dashed,thick,domain=-9:-35,scale=-3] plot ({0.95*cos(\x)}, {0.95*sin(\x)});

\draw [->,dashed,thick,domain=-81:-54,scale=-3] plot ({1.05*cos(\x)}, {1.05*sin(\x)});
\draw [->,dashed,thick,domain=-54:-81,scale=-3] plot ({0.95*cos(\x)}, {0.95*sin(\x)});

\node (A1) at (0,3.35) {};
\draw[->,thick] (A1) to [out=60,in=120,looseness=15] node[above] {$\varphi^{(1)}$} (A1);

\node (A2) at (3.35,0) {};
\draw[->,thick] (A2) to [out=-30,in=30,looseness=15] node[right] {} (A2);

\node (A3) at (0,-3.35) {};
\draw[->,thick] (A3) to [out=-60,in=-120,looseness=15] node[below] {} (A3);

\node (A4) at (-3.35,0) {};
\draw[->,thick] (A4) to [out=-150,in=-210,looseness=15] node[below] {} (A4);

\node (B1) at (0.28+1.5*1.414,0.28+1.5*1.414) {};
\draw[->,thick] (B1) to [out=15,in=75,looseness=15] node[above] {$\varphi^{(2)}$} (B1);

\node (B2) at (0.28+1.5*1.414,-0.28-1.5*1.414) {};
\draw[->,thick] (B2) to [out=-30,in=-90,looseness=15] node[right] {} (B2);

\node (B3) at (-0.28-1.5*1.414,-0.28-1.5*1.414) {};
\draw[->,thick] (B3) to [out=-105,in=-165,looseness=15] node[below right] {} (B3);

\node (B4) at (-0.28-1.5*1.414,0.28+1.5*1.414) {};
\draw[->,thick] (B4) to [out=165,in=105,looseness=15] node[above] {$\varphi^{(\alpha)}$} (B4);

\node (text) {$N$ nodes};
\circledarrow{ultra thick, gray}{text}{1.35cm};

\end{tikzpicture}
\end{center}

\caption{\it The circular quiver with gauge group $SU(K)$. The nodes denote $\NN =1$ vector multiplets. The black arrows denote $\NN= 1$ chiral multiplets and $\varphi^{(\alpha)}$, $Q^{(\alpha)}$, $\widetilde Q^{(\alpha)}$ their respective lowest components.}
\label{fig:quiver}
\end{figure}
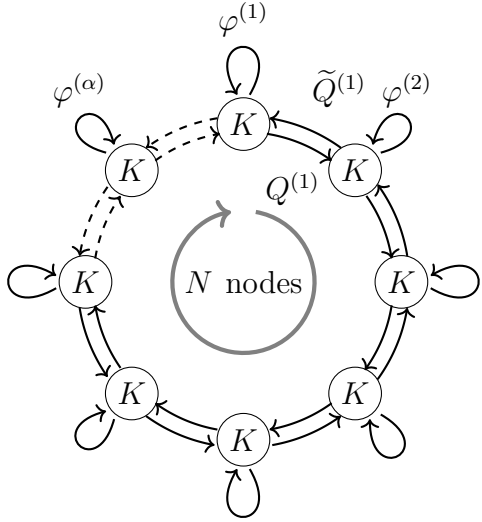

In this section we consider the more intricate behaviour of type-B conformal anomalies in superconformal $\NN = 2$ circular-quiver theories---with $N$ $SU(K)$ gauge nodes and $N$ connecting bifundamental hypermultiplets, at the `orbifold point' of equal couplings. The quiver diagram for this theory is given in Fig.~\ref{fig:quiver}. What we present here extends the calculations of Sec.~7.2 from \cite{Niarchos:2019onf} to CBOs with $\Delta = L + 1$ for $L\in \mathbb Z_{>1}$; we refer the reader to that reference for all the details of our setup. Most importantly, the results obtained here can be used to nontrivially test the conjectures of Sec.~\ref{conjectures}.

The minimal ingredients we will need from \cite{Niarchos:2019onf} are the following. Our CBOs are constructed out of $\NN = 2$ vector-multiplet adjoint scalars $\varphi^{(\alpha)}$, with $\alpha$ the node label. Following that reference, we will choose a special direction along the Higgs branch of the theory by giving the bifundamental-hypermultiplet scalars---denoted $Q^{(\alpha)}$ and $\widetilde Q^{(\alpha)}$---VEVs\footnote{This choice was necessary for the subsequent implementation of dimensional deconstruction in \cite{Niarchos:2019onf}.}
\begin{align}
  \label{eq:7}
  \langle Q^{(\alpha)} \rangle=\frac{v}{\sqrt 2} \one_{K\times K} ~, ~~
\langle \widetilde Q^{(\alpha)} \rangle =0
~.
\end{align}
This operation renders the adjoint scalars $\varphi^{(\alpha)}$ massive  and breaks the gauge symmetry down to its diagonal subgroup, $SU(K)^N\to SU(K)$.

Single-trace CBOs with $\Delta = L+1 $ comprise Casimirs of the $\varphi^{(\alpha)}$s\footnote{Once again, we are considering single-trace CBOs for simplicity but the argument goes through trivially also for multi-trace CBOs.}
\begin{align}
  \label{eq:3}
  \OO_{L+1}^{(\alpha)} \propto \Tr[(\varphi^{(\alpha)})^{L+1}]\;.
\end{align}
It is convenient to work with discrete-Fourier transformed fields
\begin{align}
  \label{eq:2}
  \OO_{L+1}^{(\alpha)} &= \frac{1}{\sqrt N} \sum_\beta \q^{\alpha\beta} \widehat \OO_{L+1}^{(\beta)}\;,\cr
  \varphi^{(\alpha)} &=  \frac{1}{\sqrt N} \sum_\beta \q^{\alpha \beta} \hat \varphi^{(\beta)}\;,
\end{align}
where $\q = e^{2\pi i/N}$ and the sum is taken over all quiver nodes. In the hatted basis the  fields $\hat\varphi^{(\alpha)}$ have mass
\begin{align}
  \label{eq:8}
  m_\alpha^2 = 2 v^2 g^2  \left( 1- \q^\alpha \right) \left( 1- \q^{-\alpha} \right)\;
\end{align}
and we can write
\begin{align}
  \label{eq:4}
  \widehat \OO^{(\alpha)}_{L+1}=
\frac{1}{\sqrt N} \sum_\beta \q^{-\alpha \beta} \Tr\left[ \left( \varphi^{(\beta)} \right)^{L+1} \right]
= \frac{1}{N^{\frac{L}{2}}} \sum_{\alpha_1,\ldots,\alpha_{L}} \Tr \left[ \left( \prod_{n=1}^{L} \hat \varphi^{(\alpha_{n})} \right) \hat \varphi^{(\alpha - \sum_{m=1}^{L} \alpha_m )} \right]
~.
\end{align}
The operators $\widehat \OO^{(\alpha)}_{L+1}$ carry $\alpha$ units of discrete-Fourier momentum. Those with $\alpha=0$ are part of the untwisted sector of the theory while those with $\alpha\neq 0$ are part of the twisted sector.

\subsection{Computation of the Tree-level Anomaly in the CFT Phase}

In the conformal phase, the leading contribution to the two-point function of such CBOs can be straightforwardly evaluated along the lines of the SCQCD example. In position space, one has
\begin{align}
  \label{eq:5}
  \langle \widehat \OO^{(\alpha)}_{L+1}(x) ~ \overline {\widehat \OO}^{(\alpha)}_{L+1}(0) \rangle
  &=\frac{1}{N^L} \sum_{\substack{\alpha_1,\ldots,\alpha_{L}\\ \alpha'_1,\ldots,\alpha'_{L}}} \left \langle \Tr \left[ \left( \prod_{n=1}^{L} \hat \varphi^{(\alpha_{n})} \right) \hat \varphi^{(\alpha - \sum_{m=1}^{L} \alpha_m )} \right](x) \times\right.\cr
  &\left.\qquad\qquad\qquad\qquad \times\Tr \left[ \left( \prod_{n=1}^{L} \hat \varphi^{(\alpha'_{n})} \right) \hat \varphi^{(\alpha' - \sum_{m=1}^{L} \alpha'_m )} \right] (0)\right \rangle\cr
  & = \frac{\CC_L^{CFT}}{(2 \pi)^{2L+2} }\frac{1}{|x|^{2L+2}}\;,
\end{align}
where like in SCQCD $\mathcal C^{CFT}_L$ denotes a colour factor, this time associated with the diagonal $SU(K)$. From this one can read out
\begin{align}
  \label{eq:6}
  G_{L+1}^{(\alpha)\rm{CFT}} =  \frac{\CC_L^{CFT}}{(2 \pi)^{2L+2} }\;.
\end{align}
We note that in the CFT phase one can use the power of supersymmetric localisation on $S^4$ to determine the two-point function for the quiver theory beyond leading order. This will be carried out in Sec.~\ref{localisation}.

\subsection{Computation of the Tree-level  Anomaly in the Higgs Phase}

We now proceed to study the type-B anomaly in the Higgs phase. The computation of the anomaly requires knowledge of the three-point function
\begin{align}
  \label{eq:11}
          \langle T(p) \widehat \OO^{(\alpha)}_{L+1}(k_1) ~ \overline {\widehat \OO}^{(\alpha)}_{L+1}(k_2) \rangle\;,
\end{align}
the evaluation of which is similar to that of the SCQCD example studied in Sec.~\ref{SCQCDHiggs}, with the following modification to the Feynman rules:
\begin{itemize}
\item The linear coupling between the trace of the energy-momentum tensor and the dilaton contributes a factor of $v p^2/2\sqrt{2}$.
\item The dilaton propagator gives a factor of $2iKN/p^2$. 
\item There is a factor of $(-i)(-\frac{\sqrt 2}{KN}g^2 v) N^{-L}\sum_{\alpha_1\neq 0} (1- \q^{\alpha_1})(1-\q^{-\alpha_1})$ from the vertex $\sigma \Tr [\varphi^{(\alpha_1)} \bar\varphi^{(\alpha_1)}]$.
\item There is an $L$-loop-momentum integral involving the scalar propagators
\begin{align}
\label{intquiv}
\mathcal{I}^{quiver}_{L,N}(k,p) := &\sum_{\alpha_1\neq 0}\sum_{\alpha_2,\ldots,\alpha_L}\int \prod_{j=1}^{L} \frac{d^4q_j}{(2\pi)^4} \frac{1}{q_1^2-m_{\alpha_1}^2}\frac{1}{(p-q_1)^2-m_{\alpha_1}^2}\times\cr
&\qquad\qquad\times\prod_{i=1}^{L-1}\frac{1}{(q_{i+1}-q_{i})^2-m_{\alpha_{i+1}}^2}\frac{1}{(k-q_L)^2-m^2_{\alpha_{L+1}}}  \;,
\end{align}
where we have used the conservation of external momenta and have defined $k$ as
\begin{equation}
    p = k_1+ k_2  \ \ \Rightarrow \ \  k_2 = p-k_1 = p-k \, \ .
  \end{equation}

\item There is a colour factor $\CC_L^{\mathbb H} = (L+1) \CC_L^{CFT}$.
\end{itemize}
Putting all these ingredients together, the three-point function contribution to the type-B anomaly along the Higgs-branch can be extracted from the relation
\begin{align}
  \label{eq:166}
 G_{L+1}^{(\alpha) (\rm dil)}= (L+1)  \CC_L^{CFT}  (-1)^{L}\frac{2^{2L-2}\Gamma(L)\Gamma(L+1)}{N^L \pi^2(2L-2)!} \sum_{\alpha_1\neq 0} m_{\alpha_1}^2\lim_{p\to 0}\lim_{k\to 0}\Big[\frac{d^{2L-2}}{dk^{2L-2}} \mathcal{I}^{quiver}_{L,N}\Big]\;.
\end{align}

At this stage we emphasise that there is a qualitative difference between the anomalies for the untwisted, $\alpha = 0$, and twisted, $\alpha\neq 0$, sectors. On the one hand, as a result of \eqref{eq:8}, the twisted CBOs involve massive degrees of freedom and are therefore lifted in the extreme IR of the theory in the broken phase. On the other hand, the untwisted CBOs involve massless fields and survive, hence leading to a nontrivial IR chiral ring. In line with our Conjecture Ib from Sec.~\ref{conjectures}, the type-B anomaly is expected to agree in the two phases, both for twisted and untwisted CBOs, but for the latter it will only do so once we take into account the contributions from the IR two-point function. Accordingly one has:
\begin{align}
  \label{eq:31}
G_{L+1}^{(0) \mathbb H} &= G_{L+1}^{(0) (\rm dil)} + G_{L+1}^{(\rm IR)}\;,\cr
G_{L+1}^{(\alpha\neq 0) \mathbb H} &= G_{L+1}^{(\alpha \neq 0) (\rm dil)}\;.
\end{align}
From now on we will focus on the anomalies of external CBOs in the untwisted sector, $\alpha = 0$, to exhibit the matching that involves the contribution of $G^{(IR)}$.

On a technical level, the main difference with SCQCD  lies in the evaluation of the integrals in \eqref{intquiv}, which are significantly more complicated than \eqref{intHIGGS} due to the presence of distinct masses. The initial steps follow those of Sec.~\ref{SCQCDHiggs} very closely. Using Feynman parametrisation the integrals can be brought to the form \eqref{eq:J}. Since in Eqs~\eqref{Mmassive}, \eqref{UKJmassive} only $J$ depends on the masses, we find
\begin{align}
  \label{eq:16}
  &  M^{quiver} = M^{massive}\;,\qquad K^{quiver} = K^{massive}\,\cr
    &  J^{quiver} = p^2x_2 + k^2 x_{L+2} - \Big(m^2_{\alpha_1}(x_1 + x_2)+\sum_{i=2}^{L+1}m^2_{\alpha_i}x_{i+1} \Big)\;,
\end{align}
which can then be plugged into \eqref{eq:result}, \eqref{eq:uf}. Employing dimensional regularisation and focussing on the $O(\epsilon^0)$ term one arrives at
\begin{align}
  \label{eq:17}
 \sum_{\alpha_1\neq 0} m_{\alpha_1}^2  \lim_{p\to 0}\lim_{k\to 0}\Big[\frac{d^{2L-2}}{dk^{2L-2}}\mathcal{I}^{quiver}_{L,N}\Big] =& \frac{(-1)^L(2L-2)!}{(4\pi)^{2L}(L-1)!}\int_0^1 \prod_{j=1}^{L+2} dx_j \ \delta\left(1-\sum_{i=1}^{L+2}x_i\right) P^{(L,N)}R^L\;,
\end{align}
where
\begin{equation}
\label{eq:QQ}
    P^{(L,N)}(x_1, \cdots, x_{L+2}) := \sum_{\alpha_1 \neq 0}\sum_{\alpha_2 \cdots \cdots \alpha_L} \frac{1}{(x_1+x_2) +m_{\alpha_1}^{-2} (\sum_{i=2}^{L+1}m_{\alpha_i}^2x_{i+1})}  \, 
\end{equation}
and
\begin{equation}
\label{eq:GG}
R^L(x_1, \cdots, x_{L+2}):= \frac{\left[(x_1+x_{2})\prod_{j=3}^{L+2}x_j\right]^{L-1}}{\mathcal{U}^{massive}(x_1,\cdots, x_{L+2})^{L+1}} \, .
\end{equation}
In this parametrisation, the difference between these integrals and those appearing in SCQCD is encoded in the $P$ factor \eqref{eq:QQ}.

In order to proceed it will be useful to introduce the following intermediate integrals 
\begin{align}
  \label{eq:19}
  \mathcal{I}_{(c,c,a_1 \cdots, a_{L})}  := \int \prod_{i=1}^{L+2}dx_i \ \delta\left(\sum_{i=1}^{L+2} x_i -1\right) \frac{1}{ c(x_1+x_2)+\sum_{i=1}^{L}a_ix_{i+2}} R^L(x_1,\cdots,x_{L+2}) \, \ ,
\end{align}
which will appear once one performs the sums over the number of nodes in $P$. Take the $L=2$, $N=2$ example for concreteness. In that case, one finds that
\begin{align}
  \label{eq:20}
 P^{(2,2)}= \frac{1}{x_1+x_2+x_3}+ \frac{1}{x_1+x_2+x_4} \, \ 
\end{align}
and therefore we can write
\begin{align}
  \label{eq:21}
 \int_0^1 \prod_{j=1}^{4} dx_j \ \delta\left(1-\sum_{i=1}^{4}x_j\right) P^{(2,2)}R^2 = \mathcal{I}_{(1,1,1,0)} + \mathcal{I}_{(1,1,0,1)}  \;.
\end{align}

Integrals of the type \eqref{eq:19} can either be evaluated outright, or obey useful identities that can then be used to compute them. For example, one finds that
\begin{align}
  \label{eq:26}
  \mathcal{I}_{(1,1,1,0)} = \mathcal{I}_{(1,1,0,1)}  = \frac{1}{4}
\end{align}
and hence
\begin{align}
  \label{eq:27}
     G_{3}^{(0) (\mathrm{dil})}= \frac{\CC_2^{CFT} }{(2\pi)^{6}}\left(1-\frac{1}{N^2}\right)\;.  
\end{align}
To avoid a long and technical detour that is needed for the general case at this stage, we have relegated all the details in the appendix. Using the results of App.~\ref{app:massive} one can show that the all $L$, all $N$ answer is 
\begin{align}
  \label{eq:22}
  \int_0^1 \prod_{j=1}^{L+2} dx_j \ \delta\left(1-\sum_{i=1}^{L+2}x_j\right) P^{(L,N)}R^L = \sum_{j=1}^{L} \frac{f_N(j+1)}{(L-j)!(j+1)!}\;,
\end{align}
where the function $f_N(x)$ is defined iteratively as
\begin{equation}
\label{eq:ff}
f_N(x) := \begin{cases}
& (N-1)^{x-1} - f_N(x-1) \ \ \textrm{for} \ \ x \geq 2\\
& 0  \ \ \textrm{for} \ \ x=1
\end{cases}\;.
\end{equation}
The series can be resummed leading to the final result
\begin{align}
  \label{eq:24}
    \int_0^1 \prod_{j=1}^{L+2} dx_j \ \delta\left(1-\sum_{i=1}^{L+2}x_i\right) P^{(L,N)}R^L  = \frac{1}{(L+1)!}(N^L -1)\;.
\end{align}
Plugging this back into \eqref{eq:166} we arrive at
\begin{align}
  \label{eq:23}
   G_{L+1}^{(0) (\mathrm{dil})}= \frac{\CC_L^{CFT} }{(2\pi)^{2L+2}}\left(1-\frac{1}{N^L}\right)\;.  
\end{align}
By performing the integrals numerically, we have also reproduced the above results for $L = 2$ up to $N=25$ and for $L=3$ up to $N=6$.\footnote{These associated computations can be found in the accompanying Mathematica file.}

As we have already mentioned in \eqref{eq:31}, for the untwisted mode, $ \alpha= 0$, there is a contribution to the anomaly from the massless fields that survive in the IR. A tree-level computation gives
\begin{align}
  \label{eq:25}
   G_{L+1}^{(\rm IR) } =  \frac{ \CC_L^{CFT} }{(2\pi)^{2L+2}N^L}\;.  
\end{align}
For the specific theory we are studying one can also extract the relevant piece in the UV tree-level computation by isolating the contributions from the massless linear combinations of the $\NN = 2$ vector multiplet. From the UV point of view, the $1/N^L$ factor originates from \eqref{eq:5}.

Upon adding the two pieces we obtain
\begin{align}
  \label{eq:32}
  G_{L+1}^{(0) \mathbb H} = G_{L+1}^{(0) (\rm dil)} + G_{L+1}^{(\rm IR)}= G_{L+1}^{(0) (\rm CFT)}\;,
\end{align}
using the definition \eqref{eq:6}. Therefore we see that for the untwisted modes the anomalies match.

For the twisted modes, the anomalies match precisely without the corresponding IR contribution \eqref{eq:31}.

\section{Type-B Anomalies for Circular Quivers from Localisation}\label{localisation}

In this section we present supersymmetric localisation computations for the type-B anomaly in the CFT phase of the $\mathcal{N}=2$ quiver gauge theory illustrated in Fig.~\ref{fig:quiver}.\footnote{For the reader interested in all the details of the calculations, we refer to \cite{Niarchos:2019onf,Mitev:2015oty} and the accompanying Mathematica file.} 
The calculations presented here extend those of \cite{Niarchos:2019onf}  beyond leading order in a weak coupling expansion, and in the $SU(2)^N$ case to also include instanton corrections. Through the anomaly-matching argument along the Higgs branch, and using the deconstruction prescription of \cite{ArkaniHamed:2001ie}, these results can provide data for type-B anomalies for the 6D (2,0) theory on $\IT^2$  \cite{Niarchos:2019onf}. 

\subsection{Preliminaries on the Partition Function on $S^4$}
We begin with the main ingredients needed for our discussion. The partition function on $S^4$ is given by the matrix integral \cite{Pestun:2007rz}
\begin{equation}
\label{eq:ZonS4}
\mathcal{Z}_{S^4}[\tau_2,\bar{\tau}_{\bar{2}};\tau_A\bar{\tau}_{\bar{A}}] = \int_{\mathfrak{t}} da \ \Delta(a) \ \mid Z(a,\tau_2,\tau_A)\mid^2 \, \ .
\end{equation}
As before, the index $\alpha=1,\cdots, N$ denotes the quiver nodes, while $a = \{a_i^{(\alpha)}\}$ labels the set of Coulomb branch parameters with $i=1,\cdots,K$. Since we are only interested in $SU(K)$ gauge groups, the Coulomb branch parameters must satisfy the constraint $\sum_{i=1}^{K}a_i^{(\alpha)} = 0 \ , \ \forall \  \alpha$.
We denote with $\tau_2 = \{\tau_2^{(\alpha)}\}$ the set of the $N$ marginal couplings, 
\begin{equation}
\tau_2^{(\alpha)} = \frac{\theta_{\alpha}}{2\pi} + i\frac{4\pi}{g^2_\alpha} \, \ , 
\end{equation}
 while we denote with $\tau_A = \{\tau_A^{(\alpha)}\}$ the set of  couplings associated with the $\Delta > 2$ CBOs.
 The integral (\ref{eq:ZonS4}) is taken over the Cartan subalgebra $\mathfrak{t}$ of the gauge group $SU(K)^N$ and $\Delta(a)$ is the corresponding Vandermonde determinant, which for the circular quiver is
  \begin{equation}
 \Delta(a) =   \Delta_\alpha(a^{(\alpha)}) = \prod_{i < j}\left(a_i^{(\alpha)}-a_j^{(\alpha)}\right)^2   \, .
\end{equation}
The function $Z(a,\tau_2,\tau_A)$ is computed via supersymmetric localisation and reads 
\begin{equation}
    Z(a,\tau,\tau_A) = Z_{cl}(a,\tau_2,\tau_A) \cdot Z_{1-loop}(a) \cdot Z_{inst}(a,\tau_2,\tau_A) \, \ .
\end{equation}
For each gauge group $SU(K)$ the classical contribution $Z_{cl}$ is
\begin{equation}
\label{eq:cl}
    Z_{cl}(a,\tau_2,\tau_A) = \textrm{exp}\left[ i \pi \tau_2 \textrm{Tr}a^2 + i\sum_{A=3}^{K}\pi^{A/2}\tau_A\textrm{Tr}a^{A}  \right] \, ,
\end{equation}
while the one-loop contribution $Z_{1-loop}$ reads
\begin{equation}
\label{eq:1loop}
    \mid Z_{1-loop}(a)\mid^2  = \frac{\prod_{ \vec{\alpha} \in \Delta^{+}( \vec{\alpha})}H^2(i  \vec{\alpha} \cdot a)}{\prod_{w \in \mathcal{R}}H(iw \cdot a)} 
    =
    \prod_{\alpha=1}^{N} \frac{\prod_{i < j}H^2(a^{\alpha}_i-a^{\beta}_j)}{\prod_{p,q=1}^KH(a_p^{\alpha}-a_{q}^{\alpha+1})}  
    \, \ ,
\end{equation}
where $H(x):=G(1+x)G(1-x)$ and $G(x)$ is the \textit{Barnes double gamma function}. The numerator contains the contributions of the vector multiplets (for the quiver in Fig.~\ref{fig:quiver}) and is given by a product over the set of the positive roots $\Delta^{+}( \vec{\alpha})$ of the Lie algebra of each gauge group $SU(K)$. The denominator contains the contributions  of the hypermultiplets and the corresponding product is taken over the weights of the representation $\mathcal{R}$ of  $SU(K) \times F$, where $F$ is the flavour symmetry group under which the hypermultiplet transforms. In our case this is the bifundamental representation  $SU(K) \times SU(K)$. Finally,  $Z_{inst}(a,\tau_2,\tau_A) $ stands for the instanton contribution.  For the case of interest in this paper, namely $SU(K)$ gauge groups, $Z_{inst}(a,\tau_2,\tau_A)$  (for $K>2$ and $\tau_A\neq0$) is currently unknown. Therefore, we will only compute instanton corrections in the case of the $SU(2)^N$ quiver theory.

\subsection{Correlation Functions of CBOs}\label{sec:corrfun}

In the CFT phase, the localisation machinery can be used to compute two-point correlation functions between a chiral $\mathcal{O}_I^{(\alpha)}(x)$ and an anti-chiral $\overline{\mathcal{O}}^{(\beta)}_{\overline{I}}(y)$ CBO with $\Delta=I$, generically associated with different nodes of the quiver,
\begin{equation}
\label{eq:G}
G_I^{\rm CFT} = G_{I}^{(\alpha,\beta)} := \langle \mathcal{O}_I^{(\alpha)}(0)\overline{\mathcal{O}}^{(\beta)}_{\overline{I}}(\infty) \rangle_{\mathbb{R}^4} \, \ .
\end{equation}
Following \cite{Gerchkovitz:2014gta,Gerchkovitz:2016gxx} the two-point correlation function on $\mathbb{R}^4$ for operators of conformal dimension $\Delta=2$ is equal to
\begin{align}
\label{eq:two2}
G^{(\alpha,\beta)}_{2} = 4^2\left[\frac{1}{\mathcal{Z}_{S^4}}\left( \partial_{\tau_2^{(\alpha)}}\partial_{\overline{\tau}_2^{(\beta)}}\mathcal{Z}_{S^4} \right) - \left( \frac{1}{\mathcal Z_{S^4}}\right)^2\partial_{\tau_2^{(\alpha)}}\mathcal{Z}_{S^4}\partial_{\overline{\tau}_2^{(\beta)}}\mathcal{Z}_{S^4}\right] \, ,
\end{align}    
while for operators with $\Delta=I>2$ we have to turn on the corresponding set of irrelevant couplings $\{\tau_I\}$ and follow the procedure outlined in \cite{Gerchkovitz:2016gxx}. For $\Delta =3$, which is the case we analyse below, the corresponding correlation functions read
\begin{align}
G_{3}^{(\alpha,\beta)} =  4^3 \frac{1}{\mathcal{Z}_{S^4}[\tau_2,\overline{\tau}_{\overline{2}}]}\partial_{\tau_3^{(\alpha)}}\partial_{\overline{\tau}_3^{(\beta)}}\mathcal{Z}_{S^4}[\tau_2,\overline{\tau}_{\overline{2}},\tau_3^{(\alpha)},\overline{\tau}_3^{(\beta)}] \Big|_{\tau_3^\alpha=\overline{\tau}_3^{(\beta)}=0} \, \ .
\end{align}

Currently it is not known how to solve the matrix integral \eqref{eq:ZonS4}. To make progress we expand the functions $H(x)$ at small $a_i^{(\alpha)}$, using the expression
\begin{equation}
    \textrm{ln} H(x) = -\sum_{n=2}^{\infty}\frac{(-1)^n}{n}\zeta(2n-1)x^{2n} \,  .
\end{equation}
In this way, order-by-order in the expansion, the integrals over the set of Coulomb-branch parameters $\{a\}$ can be performed analytically allowing us to get a perturbative expression for the partition function. 

Having performed the computation for $K=2,3,4$ we conjecture that the expression (\ref{eq:G}) for any $K$ and for $N \geq 3$ reads for the diagonal entries
\begin{align}
\label{eq:zam2diag}
 G_{2}^{(\alpha,\alpha)} &  = 2(K^2-1)\frac{g_{\alpha}^4}{(4\pi)^2} + 24(K^2-1)\Big((K^2-1)(g_{\alpha+1}^2+g_{\alpha-1}^2)-3(K^2+1)g_\alpha^2\Big)\frac{\zeta(3)g_\alpha^6}{(4\pi)^6} \cr
& - \Big[\frac{10(8K^2-12)(K^2-1)^2}{K}\Big(g_{\alpha+1}^4+g_{\alpha-1}^4+3(g_{\alpha+1}^2+g_{\alpha-1}^2)g_{\alpha}^2\Big)  g_{\alpha}^6
 \cr
&  -\frac{480}{K}(2K^2-1)(K^4-1)g_{\alpha}^{10}\Big]\frac{\zeta(5)}{(4\pi)^8} + \mathcal{O}(g_{\alpha}^{12}) 
\, ,
\end{align}
while for the non-diagonal ones
\begin{align}
\label{eq:zam2off}
 G_{2}^{(\alpha,\alpha+1)} & = 12(K^2-1)^2\frac{g_\alpha^4g_{\alpha+1}^{4} \zeta(3) }{(4\pi)^6} \ 
\cr
& + \ \frac{10(8K^2-12)(K^2-1)^2}{K}(g_{\alpha+1}^2+g_{\alpha}^2)\frac{g_{\alpha}^4g_{\alpha+1}^4\zeta(5)}{(4\pi)^8}  \ + \ \mathcal{O}(g^{12})
\end{align}
and
\begin{equation}
G_{2}^{(\alpha,\beta)} = \mathcal{O}(g^{12}) \ \ \textrm{if} \ \ \mid \alpha-\beta \mid \geq 2
\, .
\end{equation}
Moreover, based on results for $K=3,4$ we also conjecture that for any $K$ and $N \geq 3$ one has
\begin{align}
   G_{3}^{(\alpha,\alpha)} =& \frac{3(K^2-1)(K^2-4)}{K}\frac{g_{\alpha}^6}{(4\pi)^3} +\frac{54(K^2-1)(K^2-4)}{K} \times \cr
&    \times \Big((K^2-1)(g_{\alpha+1}^2+g_{\alpha-1}^2)-2(K^2+3)g_{\alpha}^2\Big)\frac{g_{\alpha}^8\zeta(3)}{(4\pi)^7} + \mathcal{O}(g^{12}) \;.
\end{align}
All other components vanish at this order.

Let us now assume $N \geq 3$. Following \cite{Niarchos:2019onf} we introduce the \textit{shift matrix} $\Omega$
\begin{equation}
\Omega_{\alpha,\beta} := \delta_{\alpha+1,\beta} \;, \ \ \ \  \alpha,\beta=1,...,N \;.
\end{equation}
We have explicitly checked for $\Delta = 2, N=10$ and $\Delta = 3, N=6$
\begin{equation}
    \Omega G_{2,3}(g_1,g_2,...,g_{N}) =  G_{2,3}(g_2,g_3,...,g_1)\Omega \, .
\end{equation}
When we move to the orbifold point, {\it i.e.} $g_\alpha \equiv g$ for $\alpha=1,...N$, $G$ and $\Omega$ are simultaneously diagonalisable. This can be implemented through a similarity transformation with $\frac{1}{\sqrt N}\mathfrak q^{\alpha \beta}$, the same matrix as the one used in \eqref{eq:2} to implement the discrete Fourier transform of the CBOs. We observe that there is always an (untwisted)  eigenvector of the form
\begin{equation}
 \textbf{v}_{un} =   (\underbrace{1,1,...,1}_{N \ \ \textrm{times}})
\end{equation}
whose eigenvalue reads
\begin{align}
 \lambda_{\textbf{v}_{un}} =&  
\frac{2}{(4\pi)^2}\left(K^2-1\right) g^4
+ \frac{144}{(4\pi)^6 }\left(K^2-1\right)\zeta(3) g^8
\cr
& \qquad  
+\frac{960}{(4\pi)^8 }\left( 3 K^3- 5 K+  \frac{2}{K}\right)\zeta(5) g^{10} + O(g^{12}) \, .
\end{align}
We observe that in the planar limit $K \rightarrow  \infty$ the eigenvalue  $\lambda_{\textbf{v}_{un}}$ is equal to the $\mathcal{N}=4$ result,  in agreement with inheritance theorems \cite{Bershadsky:1998mb,Bershadsky:1998cb}.

\subsection{Instanton Contributions for the $SU(2)^{N}$ Quiver Theory}

We will now take into account instanton corrections. We focus on  $K=2$, which is the only case for which the instanton partition function is known. For simplicity, we further restrict ourselves to the one-instanton contribution. The instanton partition function associated with each $SU(2)$ gauge node of the quiver reads 
\begin{equation}
    Z_{inst}(a^{(\alpha)},\tau^{(\alpha)}) = 1+ \frac{1}{2}e^{2\pi i\tau^{(\alpha)}}((a^{(\alpha)})^2-3) +  \cdots \, \ ,
\end{equation}
where the ellipsis denotes two and higher instanton corrections. For the $N$-noded circular quiver one has 
\begin{equation}
    Z_{inst}^{quiver}(a_\alpha,\tau_\alpha) = \prod_{\alpha=1}^{N} |Z_{inst}(a_\alpha,\tau_\alpha) |^2 \, \ .
\end{equation}
The inclusion of one-instanton corrections to the two-point functions (\ref{eq:G}) with $\Delta=2$---and by supersymmetry the Zamolodchikov metric---yields for the diagonal components
\begin{equation}
    G^{(\alpha,\alpha)}_{2} = e^{-8\pi^2/g_{\alpha}^2}\cos(\theta_{\alpha})\left(\frac{6g_{\alpha}^4}{(4\pi)^2}+\frac{12g_{\alpha}^6}{(4\pi)^4}+\frac{216g_{\alpha}^6(g_{\alpha+1}^2+g_{\alpha-1}^2-5g_{\alpha}^2)\zeta(3)}{(4\pi)^6} +\mathcal{O}(g^{10})\right)\;,
\end{equation}
while for the off-diagonal components
\begin{equation}
\begin{split}
 &  G^{(\alpha,\alpha+1)}_{2} = \frac{54g_{\alpha}^4g_{\alpha+1}^4}{(4\pi)^6}\left(e^{-8\pi^2/g_{\alpha}^2+i\theta_{\alpha}}+e^{-8\pi^2/g_{\alpha+1}^2-i\theta_{\alpha+1}}\right)\zeta(3) + \mathcal{O}\left(g^{10}e^{-8\pi^2/g^2}\right) 
\end{split}
\end{equation}
and
\begin{equation}
    G_{2}^{(\alpha,\beta)} = \mathcal{O}(g^{12}) \ \ \textrm{with} \ \ \mid \alpha-\beta \mid \geq 2 \, \ .
\end{equation}
We note that   for general $\theta_{\alpha} \neq \theta_{\beta}$ the one-instanton correction spoils the symmetry property of the Zamolodchikov metric $G_{2}^{(\alpha,\alpha+1)} \neq G_{2}^{(\alpha+1,\alpha)}$, which is now Hermitian. However, if either $\theta_{\alpha} = 0$, or one goes to the orbifold point where $g_{\alpha} \equiv g$ and $\theta_{\alpha} \equiv \theta$ the symmetric property of the Zamolodchikov metric is restored. Finally, as in the perturbative case (\ref{eq:zam2off}), the first non-trivial contribution in the perturbative expansion is proportional to $g^{8}$.

Similarly, for generic $\theta_{\alpha} \neq \theta_{\beta}$ we find that 
\begin{equation}
    \Omega G_{2}(g_1,g_2,...g_N,\theta_1,\theta_2,...,\theta_N) \neq  G_{2}(g_2,g_3,...g_1,\theta_2,\theta_3,...,\theta_1)\Omega
    \, .
\end{equation}
However, if we set $\theta_\alpha = 0$, or go to the orbifold point
\begin{align}
   \Omega G_{2}(g_1,g_2,...,g_N,0,\cdots, 0) &=G_{2}(g_2,g_3,...,g_1,0,\cdots, 0)\Omega\cr
   \Omega G_{2}(g_,\cdots,g,\theta,\cdots, \theta)&=G_{2}(g_,\cdots,g,\theta,\cdots, \theta)\Omega\;.
\end{align}
We close this discussion by mentioning that at the orbifold point, including both perturbative and one-instanton corrections and for $K=2$, the full Zamolodchikov metric always has an (untwisted) eigenvector 
\begin{equation}
 \textbf{v}_{un} =   (\underbrace{1,1,...,1}_{\ell \ \ \textrm{times}})
\end{equation}
whose eigenvalue is
\begin{equation}
\lambda_{un} =   \frac{6g^4}{(4\pi)^2}\left(1-\textrm{cos}\theta \;  e^{-8\pi^2/g^2}\left(1+\frac{2g^2}{(4\pi)^2}\right)\right) -\frac{432g^8\zeta(3)}{(4\pi)^6}\left(1-2\textrm{cos}\theta \; e^{-8\pi^2/g^2} \right) + \mathcal{O}(g^{10}) \, .
\end{equation}

\section{Anomaly Mismatch and Superconformal Manifold Holonomies}
\label{holonomy}

We will now switch gears and ask about the implications of a potential mismatch between CBO type-B anomalies on the Higgs branch,
as  raised in item $(\gamma)$ of Sec.~\ref{n2anomalies}.\footnote{Work that appeared after the publication of this paper, \cite{Schwimmer:2023nzk, Schwimmer:2024vxw}, has provided general arguments in favour of the matching of type-B
anomalies in different phases of a CFT. As a result, the discussion in this section should not be read from the perspective of potential new properties of CFTs, but from the perspective of what an assumed type-B anomaly mismatch would imply for the structure of a CFT.}

\subsection{General Arguments}\label{sec:general}

A mismatch of type-B anomalies implies that the holomorphic vector bundles of CBOs on the superconformal manifold are equipped with two symmetric rank-two tensors $G_{I\bar J}^{\rm CFT}$ and $G_{I\bar J}^{\IH}$. For concreteness, let us focus on the case of $\Delta =  2$ CBOs where, by supersymmetry, our statements translate immediately to corresponding statements about the geometry and tangent bundle of the superconformal manifold.\footnote{The discussion can be easily generalised to the holomorphic vector bundles of Coulomb-branch operators at any scaling dimension.} In that case, the first symmetric rank-two tensor is the Zamolodchikov metric, $G_{i\bar j}^{\rm CFT}$, while the second, $G_{i\bar j}^{\IH}$, is the corresponding Higgs-branch anomaly, which we will assume to be different. 

The two tensors are covariantly constant with respect to the same torsion-free connection $\nabla$ \cite{Andriolo:2022lcb}. Consequently, though different, both lead to the same Christoffel symbols. This can be achieved trivially if 
\beq
\label{holonomyaa}
G_{i\bar j}^{\IH} = C\, G_{i\bar j}^{\rm CFT} ~, ~~ C \neq 1
~,
\eeq
with $C$ a coupling-constant independent proportionality constant. A more involved possibility arises when the two tensors are genuinely different, namely when they are not proportional as in \eqref{holonomyaa}. Clearly, this can only happen when the complex dimension of the superconformal manifold is greater than one. For the rest of this discussion we assume that $G_{i\bar j}^{\IH}$ is a genuinely different covariantly constant rank-two symmetric tensor. We also assume on physical grounds that $G_{i\bar j}^{\IH}$ is globally well-defined on the superconformal manifold, based on the fact that this tensor is a Weyl anomaly after all. The existence of a second tensor $G_{i\bar j}^{\IH}$ with these properties has immediate implications for the holonomy of the superconformal manifold, as the latter must be contained within the isotropy group of $G_{i\bar j}^{\IH}$.\footnote{For 4D real Riemannian manifolds with Lorentzian signature, such restrictions have been investigated in \cite{doi:10.1063/1.529114}.}

It is known that superconformal manifolds are K\"ahler--Hodge, namely they are K\"ahler manifolds for which the flux of the K\"ahler two-form through any two-cycle is integer \cite{Gomis:2015yaa} (see also \cite{Baggio:2014ioa} for a related result). Consequently, when a superconformal manifold has complex dimension $n$, its holonomy is a subgroup of $U(n)$; for a general discussion on the holonomy theory of K\"ahler manifolds see \cite{kobayashi1963foundations,besse2007einstein}. A clean way to express this holonomy in physical terms is through the operator-state map. In that context, the holonomy of the superconformal manifold is identical to the non-abelian Berry phase of states that correspond to  exactly marginal operators in the radially quantised CFT. Since the exactly marginal operators are of the form $\Phi_i = \QQ^4 \cdot \OO_i$, $\bar \Phi_i =\bar \QQ^4 \cdot \bar \OO_i$ the Berry phase receives two contributions: one from the Berry phase of the supercharges $\QQ^\II_\alpha$, $\bar \QQ^\II_{\dot\alpha}$, and another from the Berry phase of the $\Delta = 2$ CBOs $\OO_i$, $\bar \OO_i$. The curvature of the Berry connection for the corresponding states was computed in \cite{Papadodimas:2009eu,Baggio:2014ioa,Baggio:2017aww}. Combined with techniques from supersymmetric localisation, these results can in principle be used to deduce the precise holonomy of the superconformal manifold, which in turn  must be contained within the isotropy group of $G_{i\bar j}^{\IH}$. Such a constraint can  be non-trivial when $G_{i\bar j}^{\IH}$ is not proportional to the Zamolodchikov metric. 

This observation is important, as very few general properties about the geometry of $\NN=2$ superconformal manifolds are known. In $\NN=2$ theories where  the $S^4$ partition function can be computed (e.g.\ via supersymmetric localisation), we can in principle deduce the specific form of this geometry. Typically, however, the result is very complicated and involves matrix integrals with a transseries of non-perturbative instanton corrections. The existence of $G_{i\bar j}^{\IH}$ as a second covariantly constant rank-two symmetric tensor on the conformal manifold would provide new information and constraints, which would not arise from a direct analysis of the conformal phase of the theory, but rather indirectly from an analysis of the physics of the Higgs branch.

\section{Constraints on RG Flows}\label{sec:mismatch}

Anomalies provide useful insights into the non-perturbative structure of Quantum Field Theory. In the past, chiral anomalies and type-A conformal anomalies have been used extensively to classify and constrain possible RG flows, formulate non-perturbative dualities (e.g.\ Seiberg dualities) etc. It is interesting to ask whether on can also use type-B conformal anomalies to obtain new insights into the structure of RG flows. An RG flow can be generated from a UV CFT either by turning on a VEV or by deforming the theory with a relevant operator. Let us consider first the case of VEVs.

In this paper, we discussed extensively the case of RG flows in $\NN=2$ SCFTs generated by VEVs of $\frac{1}{2}$-BPS Higgs-branch operators. The special properties of this case allowed us to uncover a rather specific picture regarding the behaviour of type-B CBO anomalies along the RG flow. We can use this picture to distill a few preliminary lessons. With matching anomalies, we found that we can use the tools available in the UV CFT description to evaluate them and deduce their exact, non-perturbative form even in the IR of the RG flow.

Although we did not uncover evidence in favor of CBO type-B anomaly mismatch on the Higgs
branch of $\mathcal N=2$ SCFTs in this paper, we would like to make the following comments. Situations where the anomalies would not match would be more complicated for computations but perhaps more intriguing, because they would seem to involve new data. Part of these data could be repackaged more compactly in terms of scheme-independent quantities that do not dependent on the conventions used in the normalisation of the CBOs.

For instance, we notice that a potential anomaly mismatch on the Higgs branch would define a non-trivial quantity of the form
\beq
\label{introea}
\delta G_{I\bar J} := G_{I\bar J}^{\rm CFT} - G_{I\bar J}^{\IH}
~.
\eeq
One of the properties of this quantity is the fact that it is covariantly constant on the superconformal manifold, $\nabla_a \delta G_{I\bar J} =0$. Then, in each subsector of the Coulomb-branch chiral ring, with fixed scaling dimension $\Delta$, we can further define a scalar
\beq
\label{introeb}
c_\Delta :=  \delta G_{I\bar J} (G^{\rm CFT})^{\bar J I} = d_\Delta - G_{I\bar J}^{\IH} (G^{\rm CFT})^{\bar J I}
~,
\eeq
where $d_\Delta$ is the dimension of the subspace of Coulomb-branch chiral superconformal primaries with scaling dimension $\Delta$. Clearly, $c_\Delta$ is a scalar independent of the normalisation of the CBOs that characterises the RG flow on the Higgs branch. This scalar is constant along the superconformal manifold, i.e.\ it is independent of the $\NN=2$ exactly marginal coupling constants  $(g^i, \bar g^i)$,
\beq
\label{introec}
\d_i c_\Delta = \d_{\bar i} c_\Delta =0 ~,
\eeq
and could therefore be deduced from a weak coupling computation (if such a computation is available).

Finally, in this paper we did not discuss situations where the RG flow is generated by relevant deformations. Therefore, we cannot use the analysis of this work to draw any immediate lessons about the fate of type-B conformal anomalies in the presence of relevant deformations. A potential approach in this direction is the study of such RG flows using a conformal compensator field along the lines of \cite{Komargodski:2011vj}. We hope to return to this aspect in a future publication.


\ack{ \bigskip We would like to thank S.~Beheshti, V.~Cort\'es and G.~Papadopoulos for helpful discussions and comments. CP is supported by the Royal Society through a University Research Fellowship, grant number UF120032, and  in part through the STFC Consolidated Grant ST/P000754/1. The work of AP and EP is partially supported by the DFG via the Emmy Noether program “Exact results in Gauge theories” and the GIF Research Grant I-1515-303./2019.}

\newpage

\begin{appendix}
  
\section{Explicit Evaluation of Massive Quiver Integrals}\label{app:massive}

In this appendix we provide a detailed derivation of the general formula Eq.~\eqref{eq:22}, used in the calculation of the massive momentum integrals of Sec.~\ref{sec:quiver}. Towards that end, some intermediate results will be useful.

\subsection{Direct Evaluation of Simple Integrals}

We first note that integrals of the type \eqref{eq:19}, in which at least one\footnote{The case in which all the $a_i=1$ was already considered in Sec.~\ref{sec:N2SCQCD}.} of the $a_i$ is equal to zero while the others are only equal to $0$ or $1$, can be computed analytically to give:
\begin{equation}
\label{eq:ana}
\mathcal{I}_{(1,1,a_1,\cdots,a_L)} =  \frac{1}{L!} \times \frac{1}{1+\sum_{i=1}^{L}a_i} \, \ .
\end{equation}
Let us prove the above assertion. We denote with $\mathcal{J}$ the set of indices $\mathcal{J}:=\{1,...,L\}$ and the formal subsets
\begin{align}
    \mathcal{J}_0 :=& \{j \in \mathcal{J} \ | \ a_j =0\} \;, \cr
    \mathcal{J}_1 :=& \{ j \in \mathcal{J} \ | \ a_j =1 \} \;.
\end{align}
As stated above we assume that $\mathcal{J} = \mathcal{J}_0 \cup \mathcal{J}_1$ and, without loss of generality, that the index $L \in \mathcal{J}_0$, that it is to say $a_L=0$. We therefore want to evaluate
\begin{align}
\mathcal{I}_{(1,1,a_1,a_2,...,a_{L-1},0)} = \int_0^1 \prod_{i=1}^{L+2} dx_i \delta\left(\sum_{i=1}^{L+2}x_i-1\right) \frac{1}{x_1+x_2+\sum_{j \in \mathcal{J}}a_jx_{j+2}}R^{L}(x_1,\cdots, x_{L+2}) \;.
\end{align}
We use the manipulations from Sec.~\ref{SCQCDHiggs} to integrate over the variable $x_2$ using the Dirac delta function, resulting in an integral over the set of variables $\{x_3,\cdots,x_{L+2},u\}$, and then also integrate over the variable $u$. This results in
\begin{align}
\label{eq:42}
   \mathcal{I}_{(1,1,a_1,a_2,...,a_{L-1},0)} &= \int_0^1 dx_3 \cdots \int_0^{1-x_3-\cdots x_{L+2}} dx_{L+2} \left(1-\sum_{i \in \mathcal{J}_1} x_i\right)\frac{ B(x_3,\cdots,x_{L+2})}{1-\sum_{j \in \mathcal{J}_0}x_j}\cr
    &=\frac{1}{L!} - \sum_{i \in \mathcal{J}_1} \int_0^1 dx_3 \cdots \int_0^{1-x_3-\cdots x_{L+2}} dx_{L+2} \frac{x_i B(x_3,\cdots,x_{L+2})}{1-\sum_{j \in \mathcal{J}_0}x_j}    \cr
& = \frac{1}{L!} - \mid \mathcal{J}_1 \mid \int_{0}^{1} \prod_{i=1}^{L+1}dx_i \ \delta\left(\sum_{k=1}^{L+1}x_k-1\right) \frac{x_1 \tilde{B}(x_1,\cdots,x_{L+1})}{1-\sum_{j\in \mathcal{J}_0}x_j} \;,
\end{align}
where $|\mathcal{J}_1|$ denotes the cardinality of the set $\mathcal{J}_1$. The function $\tilde{B}(x_1,\cdots,x_{L+1})$ is  invariant under the permutation of its arguments and is defined as
\begin{align}
  \label{eq:28}
\tilde{B}(x_1,\cdots,x_{L+1}):=  \frac{(\prod_{i = 1}^{L+1}x_i)^{L-1}}{\mathcal U^{massless}(x_j)^{L+1}}\;,
\end{align}
where $\mathcal U^{massless}(x_j)$ is given  in \eqref{Umassless}. In going from the second to the third line of \eqref{eq:42} we introduce a redundant delta function for each of the integrals and observe that all terms in the sum are equal after a relabelling of the integration variables. On the other hand, we can also introduce a redundant Dirac delta in the first line of \eqref{eq:42}, which leads to
\begin{equation}
    \mathcal{I}_{(1,1,a_1,a_2,...,a_{L-1},0)} = \int_{0}^{1} \prod_{i=1}^{L+1}dx_i \ \delta\left(\sum_{k=1}^{L+1}x_k-1\right) \frac{x_{L+1} \tilde{B}(x_1,\cdots,x_{L+1})}{1-\sum_{j\in \mathcal{J}_0}x_j} \;.
\end{equation}
Combining \eqref{eq:42} and \eqref{eq:28} we arrive at
\begin{equation}
    \mathcal{I}_{(1,1,a_1,a_2,...,a_{L-1},0)} = \frac{1}{L!} - |\mathcal{J}_1|\times \mathcal{I}_{(1,1,a_1,a_2,...,a_{L-1},0)}
\end{equation}
therefore recovering (\ref{eq:ana}), since $|\mathcal{J}_1|=\sum_{i=1}^{L}a_i$. 

\subsection{Useful Integral Identities}\label{app:genericL}

For all $a_1,a_2,\cdots,a_L \in \mathbb{C} \setminus \{0\} $, one can find  $L-1$ independent identities involving the $\mathcal I $ integrals:
\begin{align}
\label{eq:idLg1}
& \mathcal{I}_{(1,1,a_1,a_2,0,\cdots,0)} + \mathcal{I}_{(1,1,a_1^{-1}\;,a_2a_1^{-1}\;,0,\cdots,0)}  + \mathcal{I}_{(1,1,a_1a_{2}^{-1}\;,a_2^{-1}\;,0,\cdots,0)} = \frac{1}{L!} \;,\\
& \mathcal{I}_{(1,1,a_1,a_2,a_3,0,\cdots,0)} + \mathcal{I}_{(1,1,a_1^{-1}\;,a_2a_1^{-1}\;,a_3a_1^{-1}\;,0,\cdots,0)}  + \nonumber \\
&\qquad+ \mathcal{I}_{(1,1,a_1a_{2}^{-1}\;,a_2^{-1}\;,a_3a_2^{-1}\;,0,\cdots,0)} +  \mathcal{I}_{(1,1,a_1a_{3}^{-1}\;,a_2a_3^{-1},a_3^{-1}\;,0,\cdots,0)} = \frac{1}{L!}\;, \\
&  \ \ \  \ \ \ \ \ \ \ \ \ \ \  \ \ \ \  \ \ \ \  \ \ \ \ \ \ \ \ \ \ \  \ \  \ \   \ \vdots \nonumber \\
\label{eq:idLgLL}
& \mathcal{I}_{(1,1,a_1,\cdots,a_L)} + \mathcal{I}_{(1,1,a_1^{-1}\;,a_2a_1^{-1}\;,\cdots,a_La_1^{-1})} + \ \cdots \ + \mathcal{I}_{(1,1,a_1a_{L}^{-1}\;,a_2a_L^{-1}\;,\cdots,a_{L}^{-1})} = \frac{1}{L!}\;.
\end{align}

For concreteness, we consider the case with $L=2$. In this case there is  only one independent identity, namely
\begin{align}
\label{eq:identityL2}
\mathcal{I}_{(1,1,a,b)} + \mathcal{I}_{(1,1,ab^{-1},b^{-1})}+\mathcal{I}_{(1,1,a^{-1},ba^{-1})} = \frac{1}{2} \ \ \ \ \forall \ \ a,b \in \mathbb{C} \setminus \{0\}\;.
\end{align}
One can see that this identity is satisfied by considering
\begin{align}
 \mathcal{I}_{(1,1,a^{-1},ba^{-1})} &= \int_0^1 dx_1 \int_0^{1-x_1} dx_2 \frac{(1-x_1-x_2)B(x_1,x_2)}{1+x_1(a^{-1}-1)+x_2(ba^{-1}-1)}  \cr
&= \int_0^1 \prod_{i=1}^3 dx_i \delta\left(\sum_{i=1}^3x_i-1\right) \frac{x_3\widetilde{B}(x_1,x_2,x_3)}{1+x_1(a^{-1}-1)+x_2(ba^{-1}-1)} \cr
&= \int_0^1 \prod_{i=1}^3 dx_i \delta\left(\sum_{i=1}^3x_i-1\right) \frac{x_3\widetilde{B}(x_1,x_2,x_3)}{x_3+x_1a^{-1}+x_2ba^{-1}}\;. \label{eq:11a}
\end{align}
We then take
\begin{align}
 \mathcal{I}_{(1,1,ab^{-1},b^{-1})} &= \int_0^1 dx_1 \int_0^{1-x_1} dx_2 \frac{(1-x_1-x_2)B(x_1,x_2)}{1+x_1(ab^{-1}-1)+x_2(b^{-1}-1)} \cr
&= \int_0^1 \prod_{i=1}^3 dx_i \delta\left(\sum_{i=1}^3x_i-1\right) \frac{x_3\widetilde{B}(x_1,x_2,x_3)}{1+x_1(ab^{-1}-1)+x_2(b^{-1}-1)} \cr
& =\int_0^1 \prod_{i=1}^3 dx_i \delta\left(\sum_{i=1}^3x_i-1\right) \frac{x_3\widetilde{B}(x_1,x_2,x_3)}{x_3+x_1ab^{-1}+x_2b^{-1}}\label{eq:22a}
\end{align}
and
\begin{align}
 \mathcal{I}_{(1,1,a,b)} &= \int_0^1 dx_1 \int_0^{1-x_1}dx_2 \frac{1-x_1-x_2}{1+(a-1)x_1+(b-1)x_2}B(x_1,x_2)  \cr
 &= \frac{1}{2} -a\int_0^1 \prod_{i=1}^{3} dx_i \ \delta\left(\sum_{i=1}^3x_i-1\right) \frac{x_1\widetilde{B}(x_1,x_2,x_3)}{1+x_1(a-1)+x_2(b-1)} \cr
 & -b\int_0^1 \prod_{i=1}^{3} dx_i \ \delta\left(\sum_{i=1}^3x_i-1\right) \frac{x_2\widetilde{B}(x_1,x_2,x_3)}{1+x_1(a-1)+x_2(b-1)}  \cr
&= \frac{1}{2} -  \int_0^1 \prod_{i=1}^{3} dx_i \ \delta\left(\sum_{i=1}^3x_i-1\right) \frac{x_1\widetilde{B}(x_1,x_2,x_3)}{x_1+ba^{-1}x_2+a^{-1}x_3} \cr
& -\int_0^1 \prod_{i=1}^{3} dx_i \ \delta\left(\sum_{i=1}^3x_i-1\right) \frac{x_2\widetilde{B}(x_1,x_2,x_3)}{ab^{-1}x_1 +x_2 +b^{-1}x_3} \cr
&= \frac{1}{2} -\mathcal{I}_{(1,1,ab^{-1},b^{-1})} - \mathcal{I}_{(1,1,a^{-1},ba^{-1})}\;,
\end{align}
where in the last line we made use of the equations (\ref{eq:11a}) and (\ref{eq:22a}), to recover (\ref{eq:identityL2}).

For generic $L$, all identities \eqref{eq:idLg1}-\eqref{eq:idLgLL} can be proved using manipulations of this kind.

\subsection{All $L, N$ Integrals for the Circular Quiver}

We are now in a position to consider the integral on the LHS of \eqref{eq:22} for general values of $L$ and $N$. Based on the following result (see \ref{eq:ana})
\begin{equation}
\label{eq:intLL}
    \mathcal{I}_{(1,1,1,0,\cdots,0)} = \frac{1}{L!} \times \frac{1}{2} \, \ ,
\end{equation}
and on the $L-1$ independent identities \eqref{eq:idLg1}-\eqref{eq:idLgLL}, we can write
\begin{equation}
\label{eq:generalexp}
  \int_0^1 \prod_{j=1}^{L+2} dx_j \ \delta\left(1-\sum_{i=1}^{L+2}x_i\right) P^{(L,N)}R^L  = \sum_{i=1}^L \frac{1}{L!} \times \frac{c^{(L)}_i}{(i+1)}\;.
\end{equation}
Let us justify this expression. The coefficients $c_{i}^{(L)}$ are combinatorial factors that will be discussed below and carry the $N$-dependence. The numerical denominator $(i+1)L!$ originates from the analytic evaluation of the various integrals that appear once one expands $P^{(L,N)}$, defined in \eqref{eq:QQ}.

We first turn our attention to the latter by considering each contribution in the sum. The term $i=1$  accounts for integrals of the type $\mathcal{I}_{(1,1,1,0,\cdots,0)}$, while terms with $i \geq 2$ for integrals satisfying the set of identities (\ref{eq:idLg1})-(\ref{eq:idLgLL}). It can be shown that, given a generic integral $ \mathcal{I}_{(1,1,a_1,\cdots,a_L)}$, all the terms on the LHS of the identity that they participate in, (\ref{eq:idLg1})-(\ref{eq:idLgLL}), will appear as a result of the expansion of $P^{(L,N)}$. For the purposes of efficiently organising the calculation, we can assign an `effective contribution' to these terms by dividing the RHS of (\ref{eq:idLg1})-(\ref{eq:idLgLL}), namely $L!$, by the total number of terms present in a given identity, namely $i+1$.

We next consider the set of coefficients $c_i^{(L)}$ with $i=1,\cdots,L$. We claim that these can be written as 
\begin{equation}
\label{eq:cc}
    c_i^{(L)}(N) := {{L}\choose{L-i}}\times g_i(N) \, \ ,
\end{equation}
where the prefactor accounts for the equivalent ways in which one can permute $L-i$ zeros in the $L$-tuple\footnote{We remind the reader that that all these configurations lead to the same result due to the symmetry of $R^{L}(x_1,\cdots,x_{L+2})$ under permutation of its arguments.}
\begin{equation}
    (a_1,\cdots,a_{i},\underbrace{0,\cdots,0}_{L-i \ \ \textrm{times}})\;.
\end{equation}
The $N$-dependence is completely encoded in 
\begin{align}
\label{eq:gg}
g_i(N) := \begin{cases}
& N-1  \ \ \ \ \textrm{for} \ \ i=1 \\
& ((N-1)^{i-1} -f_N(i-1))(N-2) +f_N(i-1)(N-1) \ \ \ \ \textrm{for} \ \ i \geq 2
\end{cases}\;,
\end{align}
where one has iteratively 
\begin{equation}
\label{eq:ff}
f_N(x) := \begin{cases}
& (N-1)^{x-1} - f_N(x-1) \ \ \textrm{for} \ \ x \geq 2\\
& 0  \ \ \textrm{for} \ \ x=1
\end{cases}\;.
\end{equation}
The function $g_{i}(N)$ keeps track of how many terms of the given type are generated from the sums over the set $\{\alpha_1,\cdots \alpha_L\}$ in \eqref{eq:QQ}, with the index $i$ labelling how many of the coefficients $a_i$ are different from zero.

Let us justify this expression. For $i=1$ we can take $a_L \neq 0$ without loss of generality. This translates into $\alpha_2= \cdots = \alpha_{L}=0$ and only $\alpha_1 \neq 0$. Therefore there are $N-1$ terms of this type, corresponding to the range of $\alpha_1$.
 
 For $i=2$ we assume without loss of generality that $a_1 \neq 0$ and $a_L \neq 0$. This implies that contributions of this type originate from terms in the sum with $\alpha_3= \cdots =\alpha_L =0$ while $\alpha_2 \neq 0$ and $\alpha_1+\alpha_2 \neq 0$. Therefore, the number of such terms is $(N-1)$---corresponding to the range of values of $\alpha_2$---times $(N-2)$ terms---corresponding to the range of allowed values of $\alpha_1$; note that this range has been reduced by one due to the further constraint $\alpha_1 \neq -\alpha_2 $. This leads to the factor $(N-1)(N-2)$.
 
 For $i\geq3$ one encounters an additional subtlety, which we illustrate for the case of $i=3$. Without loss of generality we select $a_1$, $a_2$ and $a_L$ not equal to zero. Such terms come from having $\alpha_4= \cdots =\alpha_L =0$, so that  we are summing only over $\{\alpha_1,\alpha_2,\alpha_3\}$ while imposing the conditions $\alpha_1 \neq 0 $, $\alpha_2 \neq 0$ and $\alpha_3 \neq 0 $ and $\alpha_1 \neq -\alpha_2-\alpha_3$. Now there are two possibilities. If $\alpha_2 +\alpha_3 =0$ then $\alpha_1$ assumes $N-1$ values. Or if  $\alpha_2 +\alpha_3 \neq 0$ then $\alpha_1$ can only assume $N-2$ values. We observe that $\alpha_2 +\alpha_3 =0$ only $N-1$ times; this is counted by the function $f_N(2)$ defined in (\ref{eq:ff}). Therefore the number of times for which $\alpha_2 +\alpha_3 \neq 0$ is given by the total number of allowed combinations $(N-1)^2$ minus $N-1$. Putting everything together we arrive at the following factor for $i=3$
 \begin{equation}
     \underbrace{((N-1)^2-(N-1))}_{\textrm{Contribution for} \ \alpha_2 +\alpha_3 \neq 0} \times \underbrace{(N-2)}_{\alpha_1 \ \textrm{range}} +\underbrace{(N-1)}_{\substack{\text{Contribution for} \\ \text{$\alpha_2 +\alpha_3 = 0$}}} \times \underbrace{(N-1)}_{\alpha_1 \ \textrm{range}}\;.
 \end{equation}
 The same structures appear for higher values of $i$. Each time we have to evaluate how many times the combinations $\alpha_2+\cdots+\alpha_{i} = 0$. This is accounted for by the function $f_N(x)$ in equation (\ref{eq:ff}).

With this understanding relating to the origin of \eqref{eq:gg}, we can now further simplify it. After repeated use of the definition of $f_N(x)$, we observe that 
\begin{align}
 g_{i}(N) &= ((N-1)^{i-1}-f_N(i-1))(N-2) + f_N(i-1)(N-1)  \cr
& =f_N(i)(N-2) + f_N(i-1)(N-1) = (N-1)(f_N(i) + f_N(i-1)) - f_N(i) \cr
&= (N-1)(N-1)^{i-1} -f_N(i) = (N-1)^i -f_N(i) = f_N(i+1)\;.
\end{align}
We can finally use this to obtain
\begin{align}
  \label{eq:29}
    \int_0^1 \prod_{j=1}^{L+2} dx_j \ \delta\left(1-\sum_{i=1}^{L+2}x_i\right) P^{(L,N)}R^L  = \sum_{i=1}^L \frac{f_N(i+1)}{ (L-1)!(i+1)!} \;,
\end{align}
as advertised.

\end{appendix}



\bibliography{trphiL}

\end{document}